\documentclass[aps,prl,reprint,longbibliography,noeprint,superscriptaddress]{revtex4-2}
\usepackage{amsfonts,amssymb,amsmath,bm,graphicx,xcolor}
\usepackage[colorlinks=true,citecolor=blue,linkcolor=blue,urlcolor=blue]{hyperref}
\allowdisplaybreaks[4]

\DeclareMathOperator{\im}{Im}
\DeclareMathOperator{\sgn}{sgn}
\DeclareMathOperator{\tr}{tr}
\newcommand{\hb}[1]{\hat{\bm{#1}}}
\newcommand{\hc}[1]{\hat{\mathcal{#1}}}
\begin{document}
\title{Universal scaling of the spin Hall effect}
\author{Atsuo Shitade}
\affiliation{Institute of Scientific and Industrial Research, The University of Osaka, Ibaraki, Osaka 567-0047, Japan}
\author{Naoto Nagaosa}
\affiliation{RIKEN Center for Emergent Matter Science (CEMS), Wako, Saitama 351-0198, Japan}
\affiliation{Fundamental Quantum Science Program (FQSP), TRIP Headquarters, RIKEN, Wako, Saitama 351-0198, Japan}
\date{\today}
\begin{abstract}
  We study the spin Hall (SH) effect for the Dirac electrons
  in terms of the spin and magnetic-moment accumulation coefficients $-\Gamma^{00} g_{s(m)z}^{\phantom{s(m)z} xy}$.
  We take short-range nonmagnetic impurities into account within the self-consistent $T$-matrix approximation.
  Similarly to the universal scaling for the anomalous Hall (AH) effect,
  we find three distinct regimes by changing the electric conductivity $\sigma^{yy}$;
  the superclean regime with $-\Gamma^{00} g_{s(m)z}^{\phantom{s(m)z} xy} \propto \sigma^{yy}$ owing to the skew scattering,
  moderately dirty regime with almost constant $-\Gamma^{00} g_{s(m)z}^{\phantom{s(m)z} xy}$,
  and dirty regime with a scaling relation $-\Gamma^{00} g_{s(m)z}^{\phantom{s(m)z} xy} \propto (\sigma^{yy})^{0.6}$
  whose exponent differs from that of the AH conductivity.
  Our results construct a unified theory of the SH effect without any ambiguity of spin current.
\end{abstract}
\maketitle
\textit{Introduction.}
It has been almost 150 years since the celebrated discovery of the Hall effect~\cite{Hall1879}.
In this phenomenon, when electric and magnetic fields are applied to nonmagnetic metals perpendicular to each other,
the charge current flows in the binormal direction owing to the Lorentz force.
Soon after, a much larger response was discovered in ferromagnetic metals,
which is now called the anomalous Hall (AH) effect~\cite{RevModPhys.82.1539}.
Going back to nonmagnetic metals, the spin Hall (SH) effect occurs without a magnetic field~\cite{RevModPhys.87.1213},
in which spin current, instead of the charge current, flows perpendicular to an electric field~\cite{PhysRevLett.83.1834}
and turns into the spin accumulation at the boundaries~\cite{Dyakonov1971}.
A long history of research has revealed that both multiband and impurity effects play crucial roles in the AH and SH effects.

The microscopic origins of the AH effect are classified into the intrinsic and extrinsic mechanisms.
The former is the anomalous velocity~\cite{PhysRev.95.1154}, or the Berry curvature~\cite{RevModPhys.82.1959},
generated by the combination of the magnetization and spin-orbit coupling.
The latter comes from impurities
and is further classified into the skew scattering~\cite{Smit1955877,Smit195839} and side jump~\cite{PhysRevB.2.4559,PhysRevB.5.1862}.
A detailed study of the ferromagnetic Rashba model within the self-consistent $T$-matrix approximation
demonstrated the so-called ``universal scaling'' of the AH effect~\cite{PhysRevLett.97.126602,PhysRevB.77.165103};
the superclean regime where the AH conductivity $\sigma^{xy}$ is proportional to the longitudinal $\sigma^{yy}$ owing to the skew scattering,
moderately dirty regime where $\sigma^{xy}$ is almost constant owing to the intrinsic mechanism and side jump,
and dirty regime with $\sigma^{xy} \propto (\sigma^{yy})^{1.6}$.

It is widely accepted that the SH effect is a spin counterpart of the AH effect and originates from the same mechanisms.
The early proposals focused on the extrinsic mechanisms~\cite{Dyakonov1971,PhysRevLett.83.1834},
while the subsequent theoretical works on the intrinsic mechanism~\cite{Murakami1348,PhysRevLett.92.126603}
motivated experimental observations~\cite{Kato1910,PhysRevLett.94.047204}.
The crossover between the superclean and moderately dirty regimes was theoretically studied
based on continuum models~\cite{JPSJ.86.094704}
and using the \textit{ab initio} Korringa-Kohn-Rostoker method within the coherent potential
approximation~\cite{PhysRevLett.106.056601,PhysRevB.92.041101,PhysRevB.92.235142,PhysRevMaterials.8.015003,PhysRevB.110.184417}.
Experimentally, the moderately dirty regime has been explored
intensively~\cite{PhysRevB.94.060412,PhysRevLett.117.167204,PhysRevB.98.060410,PhysRevB.103.174427},
while the SH conductivity $\sigma_{sz}^{\phantom{sz} xy} \propto (\sigma^{yy})^{0.8}$ was reported
in the dirty regime~\cite{10.1038/s42005-021-00791-1}.
Further theoretical research is needed for the impurity effects on the SH effect.

In the presence of the spin-orbit coupling, spin is not conserved, and spin current is not uniquely defined.
Most previous \textit{ab initio} calculations
of the intrinsic SH conductivity~\cite{PhysRevLett.94.226601,PhysRevLett.95.156601,PhysRevLett.100.096401},
including the commonly used Wannier interpolation~\cite{PhysRevB.98.214402,PhysRevB.99.235113},
relied on the conventional spin current $\hat{J}_{sa}^{\phantom{sa} i} = \{\hat{s}_{a}, \hat{v}^{i}\} / 2$,
in which $\hat{s}_{a}$ and $\hat{v}^{i}$ are the spin and velocity operators, respectively.
However, this choice suffers from some critical problems
such as the nonzero uniform expectation value in equilibrium~\cite{PhysRevB.68.241315}.
Another choice is the conserved spin current~\cite{PhysRevLett.96.076604,Ma2024},
which consists of the conventional spin current and spin torque dipole moment.
This choice has some desirable properties such as the magentization spin current in equilibrium
by taking the spin torque quadrupole moment into account~\cite{PhysRevB.104.L241411}, but it remains unclear if measurable in experiments.
Some relativistic \textit{ab initio} calculations~\cite{PhysRevLett.106.056601,PhysRevB.92.041101,PhysRevB.92.235142,PhysRevMaterials.8.015003,PhysRevB.110.184417},
which include both the intrinsic and extrinsic contributions, used the conventional current
of the Bargmann-Wigner magnetic moment~\cite{pnas.34.5.211,RevModPhys.33.343,PhysRevB.76.012408}.
Interestingly, this magnetic moment is conserved in the absence of an electric or magnetic field, but not in the presence.
Hence, the problem of ambiguity is not resolved.
On the other hand, spin is well defined,
and its accumulation at the boundaries has been experimentally observed~\cite{Kato1910,PhysRevLett.94.047204}.
Spin, rather than spin current, should be the primary object in theories of the SH effect.

Recently, one of the authors proposed an alternative indicator of the SH effect called the spin accumulation coefficient (SAC),
namely, the linear response $\langle \Delta \hat{s}_{a} \rangle = g_{sa}^{\phantom{sa} ij} \partial_{x^{i}} E_{j}$ of spin
to the gradient of an electric field $E_{j}$~\cite{PhysRevB.98.174422,PhysRevB.105.L201202,PhysRevB.106.045203,Shitade2025}.
This coefficient $g_{sa}^{\phantom{sa} ij}$ characterizes the spin accumulation at the boundaries caused by the SH effect
but can be evaluated as a bulk property.
Note that the effect of diffusion is neglected.
Since the field gradient has $\delta$-function peaks at the boundaries, the same for the induced spin density.
However, by taking diffusion into account, the exponential decay of the spin accumulation can be reproduced,
and its magnitude is determined by the SAC and spin diffusion length~\cite{Shitade2025}.
The surface details are also neglected such as the Rashba spin-orbit coupling and surface states in topological insulators~\cite{RevModPhys.82.3045},
which cause the surface Edelstein effect~\cite{Ivchenko1978,Ivchenko1989,Aronov1989,Edelstein1990233}.
The advantage of this indicator is to distinguish the bulk SH effect from such surface effects~\cite{Shitade2025}.

The SAC primarily characterizes the spin accumulation for the SH effect but can be related the SH conductivity.
Within the relaxation time approximation,
$\gamma_{sa}^{\phantom{sa} ij} = -g_{sa}^{\phantom{sa} ij} / \tau$ can be interpreted as the SH conductivity because
$\partial_{x^{i}} \mathcal{J}_{sa}^{\phantom{sa} i}
= -\langle \Delta \hat{s}_{a} \rangle / \tau
= \gamma_{sa}^{\phantom{sa} ij} \partial_{x^{i}} E_{j}$,
leading to $\mathcal{J}_{sa}^{\phantom{sa} i} = \gamma_{sa}^{\phantom{sa} ij} E_{j}$.
Nonetheless, this theory does not rely on a specific definition of spin current $\mathcal{J}_{sa}^{\phantom{sa} i}$.
An important point is that the SAC vanishes in insulators, including topological insulators~\cite{RevModPhys.82.3045},
while the SH conductivity does not in general~\cite{PhysRevLett.93.156804}.
$\gamma_{sa}^{\phantom{sa} ij}$ is expressed by Bloch wave functions~\cite{PhysRevB.105.L201202},
and its formula has been implemented into \textit{ab initio} calculations~\cite{Shitade2025}.
Furthermore, the SAC for the Rashba~\cite{PhysRevB.105.L201202} and Luttinger models~\cite{PhysRevB.106.045203} was computed
within the first-order Born approximation.

In this paper, we study the impurity effects on the SAC
for the Dirac electrons describing low-energy physics of bismuth~\cite{WOLFF19641057,JPSJ.84.012001}.
We take short-range nonmagnetic impurities into account within the self-consistent $T$-matrix approximation,
as done in the previous literature on the AH effect~\cite{PhysRevLett.97.126602,PhysRevB.77.165103}.
In this system, we can also define the spin magnetic moment $\hat{m}_{a}$~\cite{JPSJ.83.074702}
and hence compute the magnetic-moment accumulation coefficient (MMAC)
$\langle \Delta \hat{m}_{a} \rangle = g_{ma}^{\phantom{ma} ij} \partial_{x^{i}} E_{j}$.
We find that both the SAC and MMAC have three regimes as functions of the electric conductivity.
In particular, a new scaling relation holds in the dirty regime.

\textit{Model and Method.}
In bismuth, low-energy electrons at the $L$ points are described by the anisotropic Dirac Hamiltonian~\cite{WOLFF19641057,JPSJ.84.012001}.
Here, we consider its isotropic case,
\begin{equation}
  \hat{H}(\bm{k})
  = -\hbar v \bm{k} \cdot \rho_{y} \bm{\sigma} + \Delta \rho_{z}, \label{eq:dirac_copy}
\end{equation}
in which $\rho$'s and $\sigma$'s are the Pauli matrices for the orbital and spin degrees of freedom, respectively.
The spin and spin magnetic-moment operators are expressed by $\hat{s}_{z} = (\hbar / 2) \sigma_{z}$
and $\hat{m}_{z} = -(g \mu_{\mathrm{B}} / 2) \rho_{z} \sigma_{z}$
with the $g$-factor $g$ and Bohr magneton $\mu_{\mathrm{B}}$~\cite{JPSJ.83.074702}.
$\rho_{z}$ in the magnetic moment can be understood from
the nonrelativistic approximation of the Dirac Hamiltonian.
The conventional current of spin vanishes as $\hat{J}_{sz}^{\phantom{sz} x} = \{\hat{s}_{z}, \hat{v}^{x}\} / 2 = 0$,
while that of the magnetic moment
$\hat{J}_{mz}^{\phantom{mz} x} = \{\hat{m}_{z}, \hat{v}^{x}\} / 2 = (g \mu_{\mathrm{B}} v / 2) \rho_{x} \sigma_{y}$
is nonzero.
Hence, the magnetic-moment Hall (MMH) conductivity
$\langle \Delta \hat{J}_{mz}^{\phantom{mz} x} \rangle = \sigma_{mz}^{\phantom{mz} xy} E_{y}$
was computed in the previous literature on the SH effect~\cite{JPSJ.81.093704,JPSJ.83.074702,JPSJ.86.094704}.
On the other hand, both the SAC and MMAC are nonzero as the symmetry allows.

We take short-range nonmagnetic impurities into account within the self-consistent $T$-matrix approximation
in the Supplemental Material (SM)~\cite{suppl}.
Following Ref.~\cite{JPSJ.86.094704}, we assume that the impurity potential is scalar as
\begin{equation}
  V_{\mathrm{i}}(\bm{r})
  = v_{\mathrm{i}} \sum_{\bm{R}_{\mathrm{i}}} \delta(\bm{r} - \bm{R}_{\mathrm{i}}), \label{eq:impurity}
\end{equation}
in which $\bm{R}_{\mathrm{i}}$ is the position of an impurity.
The retarded Green's function,
\begin{equation}
  \hat{G}^{\mathrm{R}}(\epsilon, \bm{k})
  = [\epsilon - \hat{H}(\bm{k}) - \hat{\Sigma}^{\mathrm{R}}(\epsilon)]^{-1}, \label{eq:green_1_copy}
\end{equation}
is self-consistently solved together with $\hat{\Sigma}^{\mathrm{R}}(\epsilon) = n_{\mathrm{i}} \hat{T}^{\mathrm{R}}(\epsilon)$
with the impurity concentration $n_{\mathrm{i}}$.
Here, we have introduced the $T$-matrix
$\hat{T}^{\mathrm{R}}(\epsilon) = [1 - v_{\mathrm{i}} \hc{G}^{\mathrm{R}}(\epsilon)]^{-1} v_{\mathrm{i}}$,
and $\hc{G}^{\mathrm{R}}(\epsilon)$ is the integrated Green's function,
\begin{equation}
  \hc{G}^{\mathrm{R}}(\epsilon)
  = \int \frac{d^{d} k}{(2 \pi)^{d}} \hat{G}^{\mathrm{R}}(\epsilon, \bm{k}), \label{eq:green_2_copy}
\end{equation}
with the spatial dimension $d = 3$.
The carrier density is not fixed as the impurity concentration and potential change
because of the presence of the impurity bound states.

The SAC can be computed from the linear response of spin $\hat{s}_{a}$ to a dynamical nonuniform electric field $E_{j}(\Omega, \bm{Q})$
in the first order with respect to wavenumber $\bm{Q}$ at frequency $\Omega = 0$.
The electric conductivity and SAC are computed as SM~\cite{suppl}
\begin{widetext}
  \begin{subequations}\begin{align}
    \sigma^{jj}
    = & -\frac{\hbar q^{2}}{2} \int \frac{d \epsilon}{2 \pi} f^{\prime}(\epsilon) \int \frac{d^{d} k}{(2 \pi)^{d}}
    \tr \{
      2 \hat{v}^{j} \hat{G}^{\mathrm{R}} \hat{V}^{j} \hat{G}^{\mathrm{A}}
      - [\hat{v}^{j} \hat{G}^{\mathrm{R}} \hat{v}^{j} \hat{G}^{\mathrm{R}} + (\mathrm{R} \rightarrow \mathrm{A})]
    \}. \label{eq:conductivity_1_copy} \\
    g_{sa}^{\phantom{sa} ij}
    = & \frac{i \hbar q}{4} \int \frac{d \epsilon}{2 \pi} f^{\prime}(\epsilon) \int \frac{d^{d} k}{(2 \pi)^{d}} \tr \{
      2 (
        \hat{S}_{a} \partial_{k_{i}} \hat{G}^{\mathrm{R}} \hat{V}^{j} \hat{G}^{\mathrm{A}}
        - \hat{S}_{a} \hat{G}^{\mathrm{R}} \hat{V}^{j} \partial_{k_{i}} \hat{G}^{\mathrm{A}}
      ) - [
        \hat{s}_{a} \partial_{k_{i}} \hat{G}^{\mathrm{R}} \hat{v}^{j} \hat{G}^{\mathrm{R}}
        - \hat{s}_{a} \hat{G}^{\mathrm{R}} \hat{v}^{j} \partial_{k_{i}} \hat{G}^{\mathrm{R}}
        + (\mathrm{R} \rightarrow \mathrm{A})
      ]
    \}, \label{spin_accumulation_1_copy}
  \end{align}\label{eq:coefficients}\end{subequations}
\end{widetext}
in which $f(\epsilon) = [e^{(\epsilon - \mu) / T} + 1]^{-1}$ is the Fermi distribution function
with the chemical potential $\mu$ and temperature $T$, and $q$ is the electron charge.
Hereafter, the arguments of $\epsilon$ and $\bm{k}$ are omitted for simplicity.
$\hat{S}_{a}$ and $\hat{V}^{j}$ are the renormalized spin and velocity vertices as
\begin{subequations}\begin{align}
  \hat{S}_{a}
  = & \hat{s}_{a} + n_{\mathrm{i}} \hat{T}^{\mathrm{A}}
  \int \frac{d^{d} k}{(2 \pi)^{d}} \hat{G}^{\mathrm{A}} \hat{S}_{a} \hat{G}^{\mathrm{R}}
  \hat{T}^{\mathrm{R}} \label{eq:spin_1_copy}, \\
  \hat{V}^{j}
  = & \hat{v}^{j} + n_{\mathrm{i}} \hat{T}^{\mathrm{R}}
  \int \frac{d^{d} k}{(2 \pi)^{d}} \hat{G}^{\mathrm{R}} \hat{V}^{j} \hat{G}^{\mathrm{A}}
  \hat{T}^{\mathrm{A}} \label{eq:velocity_2_copy},
\end{align}\label{eq:vertex}\end{subequations}
respectively.
The second terms are the vertex corrections (VCs).
The MMAC is available in the same way.
Note that the Fermi-sea term, which describes the intrinsic spin accumulation in the magnetic SH effect~\cite{PhysRevB.112.174431},
is prohibited in the present system with the time-reversal symmetry.
All the integrals over $\bm{k}$ are carried out analytically by introducing the dimensionless cutoff $\Lambda = 10^{3}$.
For comparison, we also employ the self-consistent first-order Born approximation with $\Lambda = 10^{2}$.
We focus on $\mu = \epsilon_{\mathrm{F}}$ and $T = 0$,
and the integrals over $\epsilon$ in Eq.~\eqref{eq:coefficients} are not necessary.

\begin{figure}
  \includegraphics[clip,width=0.48\textwidth]{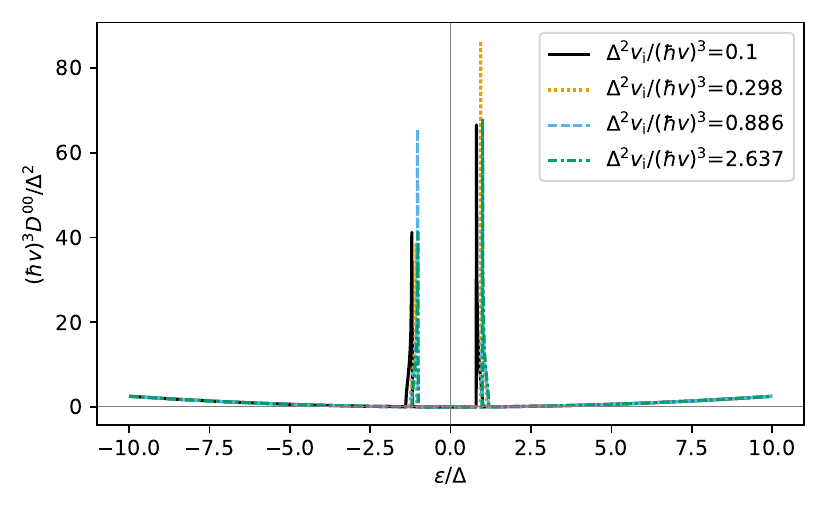}
  \caption{%
    DOS for $(\hbar v / \Delta)^{3} n_{\mathrm{i}} = 5.448$.
  } \label{fig:dos00_ni5.448}
\end{figure}

\textit{Results.}
In Fig.~\ref{fig:dos00_ni5.448}, we show the density of states (DOS)
$D^{00}(\epsilon) = -\pi^{-1} \im \mathcal{G}^{\mathrm{R} 00}(\epsilon)$
in which $\mathcal{G}^{\mathrm{R} 00}(\epsilon)$ is the $\rho_{0} \sigma_{0}$-component of Eq.~\eqref{eq:green_2_copy}.
We find two impurity bound states around
$\epsilon / \Delta = -2 \pi^{2} (\hbar v)^{3} / \Delta^{2} v_{\mathrm{i}} \Lambda \pm 1$,
which go to the band edges $\epsilon / \Delta \rightarrow \pm 1$ as $\Lambda \rightarrow \infty$.
The presence of these bound states reflects a nonperturbative nature of the approximation.

\begin{figure*}
  \includegraphics[clip,width=0.98\textwidth]{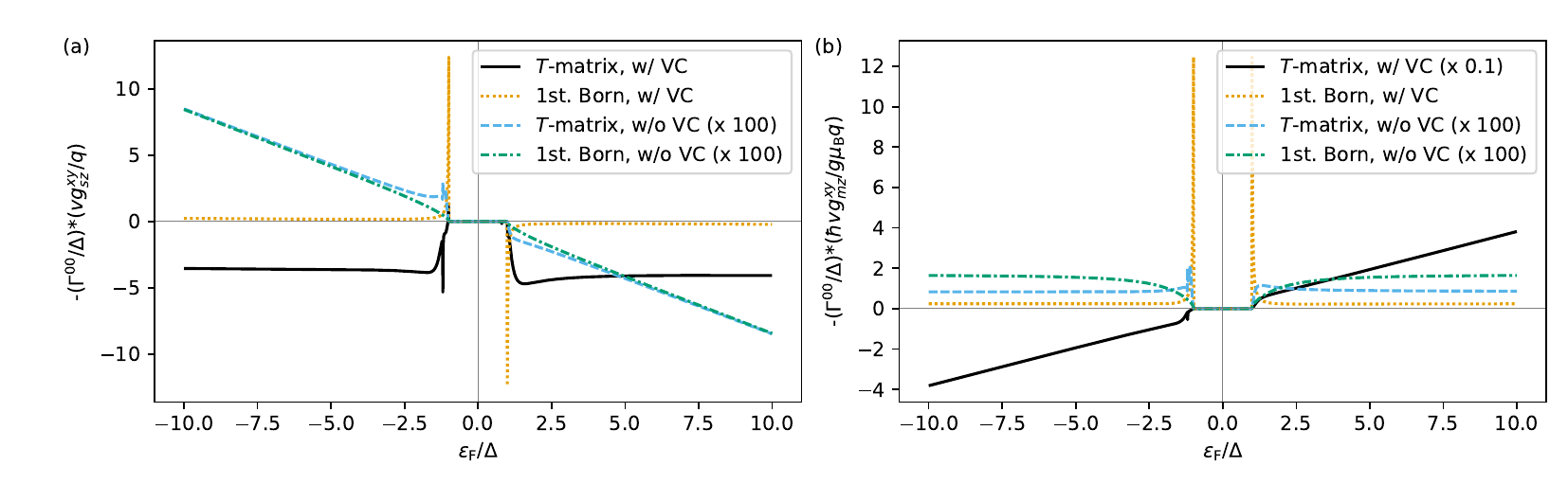}
  \caption{%
    Fermi-energy dependence of
    (a) the SAC $-\Gamma^{00} g_{sz}^{\phantom{sz} xy}$
    and (b) MMAC $-\Gamma^{00} g_{mz}^{\phantom{mz} xy}$
    within the $T$-matrix approximation with (black solid lines) and without the VCs (blue dashed lines).
    We also show the results within the first-order Born approximation with (orange dotted lines) and without the VCs (green dash-dotted lines).
    We choose $(\hbar v / \Delta)^{3} n_{\mathrm{i}} = 0.1$ and $\Delta^{2} v_{\mathrm{i}} / (\hbar v)^{3} = 0.1$.
  } \label{fig:acc_ni0.1_vi0.1}
\end{figure*}

In Fig.~\ref{fig:acc_ni0.1_vi0.1}, we show the Fermi-energy dependence of
the SAC $-\Gamma^{00} g_{sz}^{\phantom{sz} xy}$ and MMAC $-\Gamma^{00} g_{mz}^{\phantom{mz} xy}$
within the $T$-matrix and first-order Born approximations.
For comparison, we also show the results without the VCs.
Here, we have multiplied the imaginary part of the self-energy $\Gamma^{00} = -\im \Sigma^{\mathrm{R} 00}$,
which is the $\rho_{0} \sigma_{0}$-component of $\hat{\Sigma}^{\mathrm{R}}$,
so that these coefficients have the same dimensions as the SH and MMH conductivities, respectively.
Within the first-order Born approximation, with and without the VCs,
the SAC in Fig.~\ref{fig:acc_ni0.1_vi0.1}(a) is odd with respect to the Fermi energy $\epsilon_{\mathrm{F}}$,
while the MMAC in Fig.~\ref{fig:acc_ni0.1_vi0.1}(b) is even.
Within the $T$-matrix approximation with the VCs,
these coefficients are neither odd nor even owing to the skew scattering.
For $\epsilon_{\mathrm{F}} < 0$, where we have chosen $v_{\mathrm{i}} > 0$,
even their signs change by changing the impurity concentration to suppress the skew scattering.
Such behavior was already found in the MMH conductivity~\cite{JPSJ.86.094704}
and happens because the impurity potential~\eqref{eq:impurity} breaks the particle-hole symmetry of the Dirac Hamiltonian~\eqref{eq:dirac_copy}.

\begin{figure*}
  \includegraphics[clip,width=0.98\textwidth]{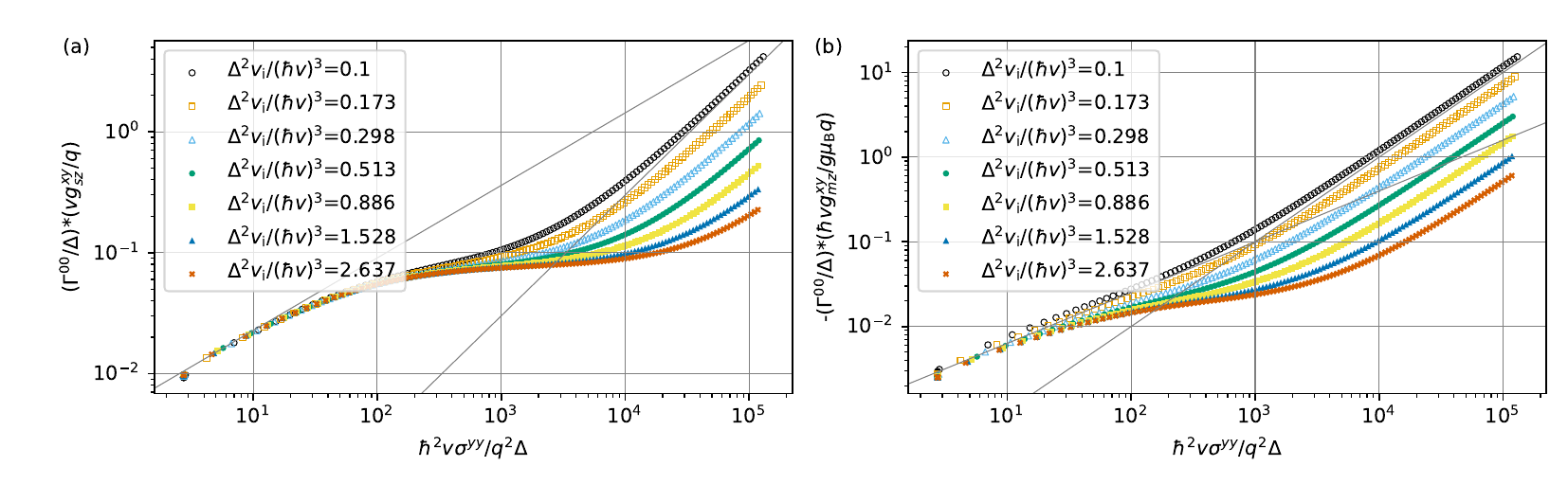}
  \caption{%
    Electric-conductivity dependence of
    (a) the SAC $\Gamma^{00} g_{sz}^{\phantom{sz} xy}$
    and MMAC $-\Gamma^{00} g_{mz}^{\phantom{mz} xy}$
    by changing the impurity concentration $(\hbar v / \Delta)^{3} n_{\mathrm{i}}$ from $10^{-1}$ to $10^{2.5}$.
    We choose $\epsilon_{\mathrm{F}} / \Delta = 3.917$.
    The slopes of two gray solid lines represent
    $\Gamma^{00} g_{sz}^{\phantom{sz} xy}, -\Gamma^{00} g_{mz}^{\phantom{mz} xy} \propto \sigma^{yy}, (\sigma^{yy})^{0.6}$.
  } \label{fig:scaling_efermi3.917}
\end{figure*}

Figure~\ref{fig:scaling_efermi3.917} shows the scaling relation between the electric conductivity and the SAC and MMAC.
We choose $\epsilon_{\mathrm{F}} > 0$ to avoid the aforementioned sign change.
Both these coefficients have three regimes similarly to the AH conductivity~\cite{PhysRevLett.97.126602,PhysRevB.77.165103}.
In the superclean regime, the skew scattering is dominant, and
$\Gamma^{00} g_{sz}^{\phantom{sz} xy}$ and $-\Gamma^{00} g_{mz}^{\phantom{mz} xy}$ are proportional to $\sigma^{yy}$.
In the moderately dirty regime, these coefficients are almost constant.
These behaviors are similar to the skew scattering and intrinsic mechanism
in terms of the MMH conductivity~\cite{JPSJ.81.093704,JPSJ.83.074702,JPSJ.86.094704}.
Finally, in the dirty regime, we find a new scaling relation
$\Gamma^{00} g_{sz}^{\phantom{sz} xy}, -\Gamma^{00} g_{mz}^{\phantom{mz} xy} \propto (\sigma^{yy})^{0.6}$.

\textit{Discussion.}
First, let us mention the moderately dirty regime.
In terms of the AH~\cite{PhysRevLett.97.126602,PhysRevB.77.165103} and SH
conductivities~\cite{JPSJ.86.094704,PhysRevLett.106.056601,PhysRevB.92.041101,PhysRevB.92.235142,PhysRevMaterials.8.015003,PhysRevB.110.184417},
the intrinsic mechanisms are dominant, which are expressed by Bloch wave functions.
On the other hand, the SAC and MMAC in nonmagnetic systems are extrinsic.
Nonetheless, if we assume the scalar self-energy and neglect the VCs,
the SAC as well as the MMAC can be computed from
Bloch energy $\epsilon_{\lambda}(\bm{k})$ and wave function $| u_{\lambda \eta}(\bm{k}) \rangle$ as SM~\cite{suppl}
\begin{align}
  g_{sa}^{\phantom{sa} ij}
  = & -\frac{q}{\hbar} \sum_{\lambda \eta \eta^{\prime}} \int \frac{d^{d} k}{(2 \pi)^{d}} \tau
  \{[s_{\lambda a}^{\phantom{\lambda a} i}(\bm{k})]_{\eta \eta^{\prime}}
  \partial_{k_{j}} \epsilon_{\lambda}(\bm{k}) \delta_{\eta^{\prime} \eta} \notag \\
  & - \epsilon^{ijb} [s_{\lambda a}(\bm{k})]_{\eta \eta^{\prime}} [m_{\lambda b}^{(\mathrm{orb})}(\bm{k})]_{\eta^{\prime} \eta}\}
  [-f^{\prime}(\epsilon_{\lambda}(\bm{k}))], \label{eq:spin_accumulation_bloch_1_copy}
\end{align}
in which $[s_{\lambda a}^{\phantom{\lambda a} i}(\bm{k})]_{\eta \eta^{\prime}},
[s_{\lambda a}(\bm{k})]_{\eta \eta^{\prime}}, [m_{\lambda b}^{(\mathrm{orb})}(\bm{k})]_{\eta^{\prime} \eta}$
are the spin magnetic quadrupole moment~\cite{PhysRevB.97.134423,PhysRevB.99.024404}, spin, and orbital magnetic moment, respectively.
Note that the previous formula~\cite{PhysRevB.105.L201202} cannot be applied to degenerate cases.
In generic multiband cases including the current one, the above assumption of the scalar self-energy is not valid,
but a possible extension is to replace the scalar relaxation time $\tau$ with the band-dependent $\tau_{\lambda}(\bm{k})$
in Eq.~\eqref{eq:spin_accumulation_bloch_1_copy}.
Thus, the SAC and MMAC are extrinsic but almost constant in the moderately dirty regime.

Next, we discuss experimental relevance of our results for bismuth.
According to the recent experiment on the spin torque efficiency in pristine and doped bismuth~\cite{PhysRevB.105.214419},
the value takes the maximum close to the Dirac point,
and its sign is independent of the carrier type.
The authors claimed that
these features are consistent with the theoretical result of the intrinsic MMH conductivity~\cite{JPSJ.81.093704,JPSJ.83.074702}.

Here, we explain these results in terms of the MMAC.
Using $\Delta = 7.65~\mathrm{meV}$ and the cyclotron mass $m_{\mathrm{c}} = \Delta / v^{2} = 0.00189 m$~\cite{PhysRevB.84.115137},
the experimental values $\epsilon_{\mathrm{F}} = 30~\mathrm{meV}$ and $\sigma^{yy} = 10^{3}~\mathrm{S/cm}$
in pristine bismuth~\cite{PhysRevB.105.214419} correspond to
$\epsilon_{\mathrm{F}} / \Delta = 3.917$ and $\hbar^{2} v \sigma^{yy} / q^{2} \Delta = 30$.
From Fig.~\ref{fig:scaling_efermi3.917}, this system is in the dirty regime, where the skew scattering is totally suppressed.
The aforementioned features are consistent with the result of the first-order Born approximation in Fig.~\ref{fig:acc_ni0.1_vi0.1}(b),
rather than Fig.~\ref{fig:acc_ni0.1_vi0.1}(a).
Furthermore, for $|\epsilon_{\mathrm{F}} / \Delta| \gg 1$, the MMAC is almost independent of the Fermi energy
and hence obeys the same scaling relation as the MMH conductivity~\cite{PhysRevB.84.115137}.
$\hbar^{2} v \sigma^{yy} / q^{2} \Delta > 900$, i.e., $\sigma^{yy} > 3 \times 10^{4}~\mathrm{S/cm}$,
is required for the superclean regime.

Let us comment on the Bargmann-Wigner magnetic moment~\cite{pnas.34.5.211,RevModPhys.33.343,PhysRevB.76.012408}.
For the Dirac Hamiltonian~\eqref{eq:dirac_copy}, another magnetic moment can be constructed as
$\hat{m}_{z}^{\mathrm{BW}} = -(g \mu_{\mathrm{B}} / 2) [\rho_{z} \sigma_{z} - (\hbar v / \Delta) \hat{k}_{z} \rho_{y}]$
with $\hat{\bm{k}} = -i \bm{\nabla}_{x} - (q / \hbar) \bm{A}(\bm{x})$,
and the corresponding current as 
$\hat{J}_{\mathrm{BW} z}^{\phantom{\mathrm{BW} z} x}
= \{\hat{m}_{z}^{\mathrm{BW}}, \hat{v}^{x}\} / 2
= (g \mu_{\mathrm{B} v} / 2) [\rho_{x} \sigma_{y} - (\hbar v / \Delta) \hat{k}_{z} \sigma_{x}]$.
This magnetic moment is conserved in the absence of an electric or magnetic field.
Even though it is not conserved under the electric field, it is interesting to study the MMH conductivity and MMAC.
However, this Hamiltonian in condensed matter physics was derived
from the low-energy $\bm{k} \cdot \bm{p}$ approximation~\cite{WOLFF19641057,JPSJ.84.012001}.
Such a magnetic moment cannot be constructed if higher-order terms are considered,
and hence we have relied on the conventional magnetic moment $\hat{m}_{z}$~\cite{JPSJ.83.074702}.

Finally, we mention applicability of our results to other systems.
We regard the Dirac model~\eqref{eq:dirac_copy} as reasonably general
since the SAC dominantly comes from the band (anti)crossing
in the moderately dirty regime as shown in Eq.~\eqref{eq:spin_accumulation_bloch_1_copy}
and even in the dirty regime where the spin magnetic quadrupole and orbital magnetic moments are smeared by strong disorder.
This mechanism is similar to the AH conductivity, which is expressed by the Berry curvature,
and is essentially described by a $2 \times 2$ model~\cite{PhysRevLett.97.126602,PhysRevB.77.165103}.
Hence, we expect that the obtained scaling relations are almost universal.
In fact, we find the same exponent $0.6$ in the dirty regime for the disordered Rashba model.
Experimentally, $\sigma_{sz}^{\phantom{sz} xy} \propto (\sigma^{yy})^{0.8}$ was reported
in the dirty regime for PtO$_{x}$ films~\cite{10.1038/s42005-021-00791-1},
though it remains unclear what definition of spin current is experimentally observed.
Further study of the impurity effect on the SH effect is neeeded for a variety of systems
both theoretically and experimentally.

\textit{Conclusion.}
We have studied the SH effect of the Dirac electrons in terms of the SAC and MMAC.
We have taken short-range nonmagnetic impurities into account within the self-consistent $T$-matrix approximation.
There are three distinct regimes as functions of the electric conductivity;
the superclean regime with $-\Gamma^{00} g_{s(m)z}^{\phantom{s(m)z} xy} \propto \sigma^{yy}$ owing to the skew scattering,
moderately dirty regime with almost constant $-\Gamma^{00} g_{s(m)z}^{\phantom{s(m)z} xy}$,
and dirty regime with the nontrivial scaling relation $-\Gamma^{00} g_{s(m)z}^{\phantom{s(m)z} xy} \propto (\sigma^{yy})^{0.6}$.
The experimentally observed carrier-density dependence of the spin torque efficiency~\cite{PhysRevB.105.214419}
is consistent with the MMAC in the dirty regime.
Our results serve as a unified theory of the SH effect without any ambiguity of spin current.

\begin{acknowledgments}
  We thank J.~Fujimoto for fruitful discussion on the $T$-matrix approximation
  and K.-J. Lee for pointing out the extension of the Bloch formula~\cite{PhysRevB.105.L201202} to degenerate cases.
  This work was supported by the Japan Society for the Promotion of Science KAKENHI Grant
  No.~JP25K07185, No.~JP24H00197, No.~JP24K00583, and No.~JP24H02231.
  N.N. was supported by RIKEN Transformative Research Innovation Platform initiative.
\end{acknowledgments}
\appendix
\begin{widetext}
\section{Self-consistent $T$-matrix approximation} \label{app:t-matrix}
We introduce the self-consistent $T$-matrix approximation.
We consider a generic form of impurities whose concentration and strength are $n_{\mathrm{i}}$ and $\hat{v}_{\mathrm{i}}(\bm{q})$,
respectively.
In this approximation, the self-energy is given by
\begin{align}
  \hat{\Sigma}^{\mathrm{R}}(\epsilon, \bm{k})
  = & n_{\mathrm{i}} \left[
    \hat{v}_{\mathrm{i}}(0)
    + \int \frac{d^{d} q_{1}}{(2 \pi)^{d}}
    \hat{v}_{\mathrm{i}}(\bm{q}_{1}) \hat{G}^{\mathrm{R}}(\epsilon, \bm{k} + \bm{q}_{1}) \hat{v}_{\mathrm{i}}(-\bm{q}_{1})
  \right. \notag \\
  & \left.
    + \int \frac{d^{d} q_{1}}{(2 \pi)^{d}} \int \frac{d^{d} q_{1}}{(2 \pi)^{d}}
    \hat{v}_{\mathrm{i}}(\bm{q}_{1}) \hat{G}^{\mathrm{R}}(\epsilon, \bm{k} + \bm{q}_{1})
    \hat{v}_{\mathrm{i}}(-\bm{q}_{1} + \bm{q}_{2}) \hat{G}^{\mathrm{R}}(\epsilon, \bm{k} + \bm{q}_{2}) \hat{v}_{\mathrm{i}}(-\bm{q}_{2})
    + \dots
  \right], \label{eq:t-matrix_1}
\end{align}
with the Green's function,
\begin{equation}
  \hat{G}^{\mathrm{R}}(\epsilon, \bm{k})
  = [\epsilon - \hat{H}(\bm{k}) - \hat{\Sigma}^{\mathrm{R}}(\epsilon, \bm{k})]^{-1}. \label{eq:green_1}
\end{equation}
Here, $d$ is the spatial dimension.
Equation~\eqref{eq:t-matrix_1} is schematically shown in Fig.~\ref{fig:diagram}(a).
In the case of short-range impurities, $\hat{v}_{\mathrm{i}}(\bm{q})$ is independent of $\bm{q}$,
and the self-energy becomes independent of $\bm{k}$ as
\begin{align}
  \hat{\Sigma}^{\mathrm{R}}(\epsilon)
  = & n_{\mathrm{i}} \left[
    \hat{v}_{\mathrm{i}}
    + \hat{v}_{\mathrm{i}} \int \frac{d^{d} q_{1}}{(2 \pi)^{d}} \hat{G}^{\mathrm{R}}(\epsilon, \bm{k} + \bm{q}_{1}) \hat{v}_{\mathrm{i}}
    + \hat{v}_{\mathrm{i}} \int \frac{d^{d} q_{1}}{(2 \pi)^{d}} \hat{G}^{\mathrm{R}}(\epsilon, \bm{k} + \bm{q}_{1})
    \hat{v}_{\mathrm{i}} \int \frac{d^{d} q_{1}}{(2 \pi)^{d}} \hat{G}^{\mathrm{R}}(\epsilon, \bm{k} + \bm{q}_{2}) \hat{v}_{\mathrm{i}}
    + \dots
  \right] \notag \\
  = & n_{\mathrm{i}} \left\{
    1 + \hat{v}_{\mathrm{i}} \hc{G}^{\mathrm{R}}(\epsilon) + [\hat{v}_{\mathrm{i}} \hc{G}^{\mathrm{R}}(\epsilon)]^{2} + \dots
  \right\} \hat{v}_{\mathrm{i}} \notag \\
  = & n_{\mathrm{i}} [1 - \hat{v}_{\mathrm{i}} \hc{G}^{\mathrm{R}}(\epsilon)]^{-1} \hat{v}_{\mathrm{i}} \notag \\
  = & n_{\mathrm{i}} \hat{T}^{\mathrm{R}}(\epsilon). \label{eq:t-matrix_2}
\end{align}
Here, we have introduced the integrated Green's function as
\begin{equation}
  \hc{G}^{\mathrm{R}}(\epsilon)
  = \int \frac{d^{d} k}{(2 \pi)^{d}} \hat{G}^{\mathrm{R}}(\epsilon, \bm{k}), \label{eq:green_2}
\end{equation}
and $\hat{T}^{\mathrm{R}}(\epsilon) = [1 - \hat{v}_{\mathrm{i}} \hc{G}^{\mathrm{R}}(\epsilon)]^{-1} \hat{v}_{\mathrm{i}}$.
The energy derivative is available from
\begin{subequations}\begin{align}
  \hc{G}^{\mathrm{R} \prime}(\epsilon)
  = & -\int \frac{d^{d} k}{(2 \pi)^{d}} [\epsilon - \hat{H}(\bm{k}) - \hat{\Sigma}^{\mathrm{R}}(\epsilon, \bm{k})]^{-1}
  [1 - \hat{\Sigma}^{\mathrm{R} \prime}(\epsilon)] [\epsilon - \hat{H}(\bm{k}) - \hat{\Sigma}^{\mathrm{R}}(\epsilon, \bm{k})]^{-1} \notag \\
  = & -\int \frac{d^{d} k}{(2 \pi)^{d}} \hat{G}^{\mathrm{R}}(\epsilon, \bm{k})
  [1 - \hat{\Sigma}^{\mathrm{R} \prime}(\epsilon)] \hat{G}^{\mathrm{R}}(\epsilon, \bm{k}), \label{eq:t-matrix_3a} \\
  \hat{T}^{\mathrm{R} \prime}(\epsilon)
  = & [1 - \hat{v}_{\mathrm{i}} \hc{G}^{\mathrm{R}}(\epsilon)]^{-1} \hat{v}_{\mathrm{i}} \hc{G}^{\mathrm{R} \prime}(\epsilon)
  [1 - \hat{v}_{\mathrm{i}} \hc{G}^{\mathrm{R}}(\epsilon)]^{-1} \hat{v}_{\mathrm{i}}
  = \hat{T}^{\mathrm{R}}(\epsilon) \hc{G}^{\mathrm{R} \prime}(\epsilon) \hat{T}^{\mathrm{R}}(\epsilon), \label{eq:t-matrix_3b}
\end{align}\label{eq:t-matrix_3}\end{subequations}
which is necessary for the SH conductivity in Eq.~\eqref{eq:spin_hall_1}.
\begin{figure*}
  \includegraphics[clip,width=0.98\textwidth]{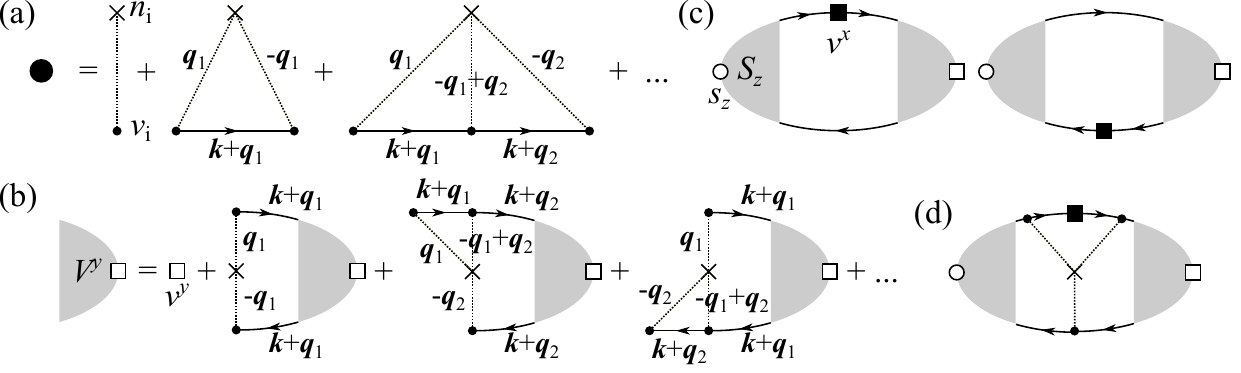}
  \caption{%
    Feynman diagrams for (a) the self-energy within the self-consistent $T$-matrix approximation,
    (b) the renormalized velocity vertex,
    (c) the spin--charge-current correlation function of the first order with respect to $Q_{x}$,
    and (d) the correation function that vanishes in the case of short-range impurities.
  } \label{fig:diagram}
\end{figure*}

We need to take VCs at $\bm{Q} = 0$ into account for the Fermi-surface terms in transport coefficients.
The renormalized velocity vertex is obtained by solving
\begin{align}
  & \hat{V}^{j}(\epsilon, \bm{k}) \notag \\
  = & \hat{v}^{j}(\bm{k})
  + n_{\mathrm{i}} \left[
    \int \frac{d^{d} q_{1}}{(2 \pi)^{d}} \hat{v}_{\mathrm{i}}(\bm{q}_{1}) \hat{G}^{\mathrm{R}}(\epsilon, \bm{k} + \bm{q}_{1})
    \hat{V}^{j}(\epsilon, \bm{k} + \bm{q}_{1}) \hat{G}^{\mathrm{A}}(\epsilon, \bm{k} + \bm{q}_{1}) \hat{v}_{\mathrm{i}}(-\bm{q}_{1})
  \right. \notag \\
    & + \int \frac{d^{d} q_{1}}{(2 \pi)^{d}} \int \frac{d^{d} q_{2}}{(2 \pi)^{d}} \hat{v}_{\mathrm{i}}(\bm{q}_{1})
    \hat{G}^{\mathrm{R}}(\epsilon, \bm{k} + \bm{q}_{1}) \hat{v}_{\mathrm{i}}(-\bm{q}_{1} + \bm{q}_{2})
    \hat{G}^{\mathrm{R}}(\epsilon, \bm{k} + \bm{q}_{2}) \hat{V}^{j}(\epsilon, \bm{k} + \bm{q}_{2})
    \hat{G}^{\mathrm{A}}(\epsilon, \bm{k} + \bm{q}_{2}) \hat{v}_{\mathrm{i}}(-\bm{q}_{2}) \notag \\
  & \left.
    + \int \frac{d^{d} q_{1}}{(2 \pi)^{d}} \int \frac{d^{d} q_{2}}{(2 \pi)^{d}} \hat{v}_{\mathrm{i}}(\bm{q}_{1})
    \hat{G}^{\mathrm{R}}(\epsilon, \bm{k} + \bm{q}_{1}) \hat{V}^{j}(\epsilon, \bm{k} + \bm{q}_{1})
    \hat{G}^{\mathrm{A}}(\epsilon, \bm{k} + \bm{q}_{1}) \hat{v}_{\mathrm{i}}(-\bm{q}_{1} + \bm{q}_{2})
    \hat{G}^{\mathrm{A}}(\epsilon, \bm{k} + \bm{q}_{2}) \hat{v}_{\mathrm{i}}(-\bm{q}_{2})
    + \dots
  \right], \label{eq:velocity_1}
\end{align}
which is schematically shown in Fig.~\ref{fig:diagram}(b).
In the case of short-range impurities, this equation is much simplified as
\begin{align}
  \hat{V}^{j}(\epsilon, \bm{k})
  = & \hat{v}^{j}(\bm{k})
  + n_{\mathrm{i}} \left[
    \hat{v}_{\mathrm{i}} \int \frac{d^{d} q_{1}}{(2 \pi)^{d}} \hat{G}^{\mathrm{R}}(\epsilon, \bm{k} + \bm{q}_{1})
    \hat{V}^{j}(\epsilon, \bm{k} + \bm{q}_{1}) \hat{G}^{\mathrm{A}}(\epsilon, \bm{k} + \bm{q}_{1}) \hat{v}_{\mathrm{i}}
  \right. \notag \\
    & + \hat{v}_{\mathrm{i}} \int \frac{d^{d} q_{1}}{(2 \pi)^{d}} \hat{G}^{\mathrm{R}}(\epsilon, \bm{k} + \bm{q}_{1}) \hat{v}_{\mathrm{i}}
    \int \frac{d^{d} q_{2}}{(2 \pi)^{d}} \hat{G}^{\mathrm{R}}(\epsilon, \bm{k} + \bm{q}_{2}) \hat{V}^{j}(\epsilon, \bm{k} + \bm{q}_{2})
    \hat{G}^{\mathrm{A}}(\epsilon, \bm{k} + \bm{q}_{2}) \hat{v}_{\mathrm{i}} \notag \\
  & \left.
    + \hat{v}_{\mathrm{i}} \int \frac{d^{d} q_{1}}{(2 \pi)^{d}} \hat{G}^{\mathrm{R}}(\epsilon, \bm{k} + \bm{q}_{1})
    \hat{V}^{j}(\epsilon, \bm{k} + \bm{q}_{1}) \hat{G}^{\mathrm{A}}(\epsilon, \bm{k} + \bm{q}_{1}) \hat{v}_{\mathrm{i}}
    \int \frac{d^{d} q_{2}}{(2 \pi)^{d}} \hat{G}^{\mathrm{A}}(\epsilon, \bm{k} + \bm{q}_{2}) \hat{v}_{\mathrm{i}}
    + \dots
  \right] \notag \\
  = & \hat{v}^{j}(\bm{k})
  + n_{\mathrm{i}} [1 + \hat{v}_{\mathrm{i}} \hc{G}^{\mathrm{R}}(\epsilon) + \dots] \hat{v}_{\mathrm{i}}
  \hc{V}^{j}(\epsilon) \hat{v}_{\mathrm{i}} [1 + \hc{G}^{\mathrm{A}}(\epsilon) \hat{v}_{\mathrm{i}} + \dots] \notag \\
  = & \hat{v}^{j}(\bm{k})
  + n_{\mathrm{i}} \hat{T}^{\mathrm{R}}(\epsilon) \hc{V}^{j}(\epsilon) \hat{T}^{\mathrm{A}}(\epsilon), \label{eq:velocity_2}
\end{align}
using Eq.~\eqref{eq:green_2} and
\begin{equation}
  \hc{V}^{j}(\epsilon)
  = \int \frac{d^{d} k}{(2 \pi)^{d}} \hat{G}^{\mathrm{R}}(\epsilon, \bm{k}) \hat{V}^{j}(\epsilon, \bm{k})
  \hat{G}^{\mathrm{A}}(\epsilon, \bm{k}). \label{eq:velocity_3}
\end{equation}
Similarly, the renormalized spin vertex is obtained by solving
\begin{equation}
  \hat{S}_{a}(\epsilon)
  = \hat{s}_{a} + n_{\mathrm{i}} \hat{T}^{\mathrm{A}}(\epsilon) \hc{S}_{a}(\epsilon) \hat{T}^{\mathrm{A}}(\epsilon), \label{eq:spin_1}
\end{equation}
with
\begin{equation}
  \hc{S}_{a}(\epsilon)
  = \int \frac{d^{d} k}{(2 \pi)^{d}} \hat{G}^{\mathrm{A}}(\epsilon, \bm{k}) \hat{S}_{a}(\epsilon)
  \hat{G}^{\mathrm{R}}(\epsilon, \bm{k}). \label{eq:spin_2}
\end{equation}

\section{Transport coefficients} \label{app:transport}
We calculate the linear response of a given operator $\hat{B}$ to a vector potential $A_{j}(\Omega, \bm{Q})$.
Using the Keldysh Green's function $\hat{G}(\epsilon, \bm{k})$, the response is expressed by
\begin{align}
  \langle \Delta \hat{B} \rangle(\Omega, \bm{Q})
  = & i q A_{j}(\Omega, \bm{Q}) \int \frac{d \epsilon}{2 \pi} \int \frac{d^{d} k}{(2 \pi)^{d}}
  \tr [\hat{B}(\bm{k}; \bm{Q}) \hat{G}(\epsilon_{+}, \bm{k}_{+}) \hat{v}^{j}(\bm{k}, \bm{Q}) \hat{G}(\epsilon_{-}, \bm{k}_{-})]^{<} \notag \\
  = & i q A_{j}(\Omega, \bm{Q}) \int \frac{d \epsilon}{2 \pi} \int \frac{d^{d} k}{(2 \pi)^{d}} \tr \{
    \hat{B}(\bm{k}; \bm{Q}) \hat{G}^{\mathrm{R}}(\epsilon_{+}, \bm{k}_{+}) \hat{v}^{j}(\bm{k}, \bm{Q})
    [\hat{G}^{\mathrm{A}}(\epsilon_{-}, \bm{k}_{-}) - \hat{G}^{\mathrm{R}}(\epsilon_{-}, \bm{k}_{-})] f(\epsilon_{-}) \notag \\
    & + \hat{B}(\bm{k}; \bm{Q}) [\hat{G}^{\mathrm{A}}(\epsilon_{+}, \bm{k}_{+}) - \hat{G}^{\mathrm{R}}(\epsilon_{+}, \bm{k}_{+})]
    \hat{v}^{j}(\bm{k}, \bm{Q}) \hat{G}^{\mathrm{A}}(\epsilon_{-}, \bm{k}_{-}) f(\epsilon_{+})
  \} \notag \\
  = & i q A_{j}(\Omega, \bm{Q}) \int \frac{d \epsilon}{2 \pi} \int \frac{d^{d} k}{(2 \pi)^{d}} \tr \{
    \hat{B}(\bm{k}; \bm{Q}) \hat{G}^{\mathrm{R}}(\epsilon_{+}, \bm{k}_{+}) \hat{v}^{j}(\bm{k}, \bm{Q})
    \hat{G}^{\mathrm{A}}(\epsilon_{-}, \bm{k}_{-}) [f(\epsilon_{-}) - f(\epsilon_{+})] \notag \\
    & - \hat{B}(\bm{k}; \bm{Q}) \hat{G}^{\mathrm{R}}(\epsilon_{+}, \bm{k}_{+}) \hat{v}^{j}(\bm{k}, \bm{Q})
    \hat{G}^{\mathrm{R}}(\epsilon_{-}, \bm{k}_{-}) f(\epsilon_{-}) \notag \\
    & + \hat{B}(\bm{k}; \bm{Q}) \hat{G}^{\mathrm{A}}(\epsilon_{+}, \bm{k}_{+}) \hat{v}^{j}(\bm{k}, \bm{Q})
    \hat{G}^{\mathrm{A}}(\epsilon_{-}, \bm{k}_{-}) f(\epsilon_{+})
  \}. \label{eq:transport_1}
\end{align}
Here, we have used
$G^{<}(\epsilon, \bm{k}) = [G^{\mathrm{A}}(\epsilon, \bm{k}) - G^{\mathrm{R}}(\epsilon, \bm{k})] f(\epsilon)$
for the lesser Green's function with $f(\epsilon) = [e^{(\epsilon - \mu) / T} + 1]^{-1}$ being the Fermi distribution function.
$\mu$ and $T$ are the chemical potential and temperature, and $q$ is the electron charge.
$\epsilon_{\pm} = \epsilon \pm \hbar \Omega / 2$ and $\bm{k}_{\pm} = \bm{k} \pm \bm{Q} / 2$,
and $\hat{B}(\bm{k}; \bm{Q}) = [\hat{B}(\bm{k}_{+}) + \hat{B}(\bm{k}_{-})] / 2$.
We expand this expression up to the first order with respect to $\Omega$ and $\bm{Q}$ as
\begin{align}
  \langle \Delta \hat{B} \rangle(\Omega, \bm{Q})
  = & -i q A_{j}(\Omega, \bm{Q}) \int \frac{d \epsilon}{2 \pi} f(\epsilon) \int \frac{d^{d} k}{(2 \pi)^{d}} \tr [
    \hat{B}(\bm{k}; \bm{Q}) \hat{G}^{\mathrm{R}}(\epsilon, \bm{k}_{+}) \hat{v}^{j}(\bm{k}, \bm{Q}) \hat{G}^{\mathrm{R}}(\epsilon, \bm{k}_{-})
    - (\mathrm{R} \rightarrow \mathrm{A})
  ] \notag \\
  & - \frac{i \hbar q}{2} \Omega A_{j}(\Omega, \bm{Q}) \int \frac{d \epsilon}{2 \pi} f(\epsilon) \int \frac{d^{d} k}{(2 \pi)^{d}} \notag \\
  & \times \tr [
    \hat{B}(\bm{k}; \bm{Q}) \partial_{\epsilon} \hat{G}^{\mathrm{R}}(\epsilon, \bm{k}_{+}) \hat{v}^{j}(\bm{k}, \bm{Q})
    \hat{G}^{\mathrm{R}}(\epsilon, \bm{k}_{-})
    - \hat{B}(\bm{k}; \bm{Q}) \hat{G}^{\mathrm{R}}(\epsilon, \bm{k}_{+}) \hat{v}^{j}(\bm{k}, \bm{Q})
    \partial_{\epsilon} \hat{G}^{\mathrm{R}}(\epsilon, \bm{k}_{-})
    - (\mathrm{R} \rightarrow \mathrm{A})
  ] \notag \\
  & - \frac{i \hbar q}{2} \Omega A_{j}(\Omega, \bm{Q}) \int \frac{d \epsilon}{2 \pi} f^{\prime}(\epsilon) \int \frac{d^{d} k}{(2 \pi)^{d}} \notag \\
  & \times \tr \{
    2 \hat{B}(\bm{k}; \bm{Q}) \hat{G}^{\mathrm{R}}(\epsilon, \bm{k}_{+}) \hat{v}^{j}(\bm{k}, \bm{Q}) \hat{G}^{\mathrm{A}}(\epsilon, \bm{k}_{-})
    - [
      \hat{B}(\bm{k}; \bm{Q}) \hat{G}^{\mathrm{R}}(\epsilon, \bm{k}_{+}) \hat{v}^{j}(\bm{k}, \bm{Q}) \hat{G}^{\mathrm{R}}(\epsilon, \bm{k}_{-})
      + (\mathrm{R} \rightarrow \mathrm{A})
    ]
  \} \notag \\
  = & -i q A_{j}(\Omega, \bm{Q}) \int \frac{d \epsilon}{2 \pi} f(\epsilon) \int \frac{d^{d} k}{(2 \pi)^{d}}
  \tr [\hat{B} \hat{G}^{\mathrm{R}} \hat{v}^{j} \hat{G}^{\mathrm{R}} - (\mathrm{R} \rightarrow \mathrm{A})] \notag \\
  & - \frac{i \hbar q}{2} \Omega A_{j}(\Omega, \bm{Q}) \int \frac{d \epsilon}{2 \pi} f(\epsilon) \int \frac{d^{d} k}{(2 \pi)^{d}} \tr [
    \hat{B} \partial_{\epsilon} \hat{G}^{\mathrm{R}} \hat{v}^{j} \hat{G}^{\mathrm{R}}
    - \hat{B} \hat{G}^{\mathrm{R}} \hat{v}^{j} \partial_{\epsilon} \hat{G}^{\mathrm{R}}
    - (\mathrm{R} \rightarrow \mathrm{A})
  ] \notag \\
  & - \frac{i \hbar q}{2} \Omega A_{j}(\Omega, \bm{Q}) \int \frac{d \epsilon}{2 \pi} f^{\prime}(\epsilon)
  \int \frac{d^{d} k}{(2 \pi)^{d}} \tr \{
    2 \hat{B} \hat{G}^{\mathrm{R}} \hat{v}^{j} \hat{G}^{\mathrm{A}}
    - [\hat{B} \hat{G}^{\mathrm{R}} \hat{v}^{j} \hat{G}^{\mathrm{R}} + (\mathrm{R} \rightarrow \mathrm{A})]
    \} \notag \\
  & - \frac{i q}{2} Q_{i} A_{j}(\Omega, \bm{Q}) \int \frac{d \epsilon}{2 \pi} f(\epsilon) \int \frac{d^{d} k}{(2 \pi)^{d}} \tr [
    \hat{B} \partial_{k_{i}} \hat{G}^{\mathrm{R}} \hat{v}^{j} \hat{G}^{\mathrm{R}}
    - \hat{B} \hat{G}^{\mathrm{R}} \hat{v}^{j} \partial_{k_{i}} \hat{G}^{\mathrm{R}}
    - (\mathrm{R} \rightarrow \mathrm{A})
  ] \notag \\
  & - \frac{i \hbar q}{4} \Omega Q_{i} A_{j}(\Omega, \bm{Q}) \int \frac{d \epsilon}{2 \pi} f(\epsilon)
  \int \frac{d^{d} k}{(2 \pi)^{d}} \notag \\
  & \times \tr [
    \hat{B} \partial_{\epsilon} \partial_{k_{i}} \hat{G}^{\mathrm{R}} \hat{v}^{j} \hat{G}^{\mathrm{R}}
    - \hat{B} \partial_{\epsilon} \hat{G}^{\mathrm{R}} \hat{v}^{j} \partial_{k_{i}} \hat{G}^{\mathrm{R}}
    + \hat{B} \hat{G}^{\mathrm{R}} \hat{v}^{j} \partial_{\epsilon} \partial_{k_{i}} \hat{G}^{\mathrm{R}}
    - \hat{B} \partial_{k_{i}} \hat{G}^{\mathrm{R}} \hat{v}^{j} \partial_{\epsilon} \hat{G}^{\mathrm{R}}
    - (\mathrm{R} \rightarrow \mathrm{A})
  ] \notag \\
  & - \frac{i \hbar q}{4} \Omega Q_{i} A_{j}(\Omega, \bm{Q}) \int \frac{d \epsilon}{2 \pi} f^{\prime}(\epsilon)
  \int \frac{d^{d} k}{(2 \pi)^{d}} \notag \\
  & \times \tr \{
    2 (
      \hat{B} \partial_{k_{i}} \hat{G}^{\mathrm{R}} \hat{v}^{j} \hat{G}^{\mathrm{A}}
      - \hat{B} \hat{G}^{\mathrm{R}} \hat{v}^{j} \partial_{k_{i}} \hat{G}^{\mathrm{A}}
    ) - [
      \hat{B} \partial_{k_{i}} \hat{G}^{\mathrm{R}} \hat{v}^{j} \hat{G}^{\mathrm{R}}
      - \hat{B} \hat{G}^{\mathrm{R}} \hat{v}^{j} \partial_{k_{i}} \hat{G}^{\mathrm{R}}
      + (\mathrm{R} \rightarrow \mathrm{A})
    ]
  \}. \label{eq:transport_2}
\end{align}
Here, we have omitted the variables $\epsilon$ and $\bm{k}$ for simplicity.
The linear response of the charge current to an electric field $E_{j}(\Omega, \bm{Q}) = i \Omega A_{j}(\Omega, \bm{Q})$
yields the electric conductivity as
\begin{equation}
  \sigma^{jj}
  = -\frac{\hbar q^{2}}{2} \int \frac{d \epsilon}{2 \pi} f^{\prime}(\epsilon) \int \frac{d^{d} k}{(2 \pi)^{d}}
  \tr \{
    2 \hat{v}^{j} \hat{G}^{\mathrm{R}} \hat{V}^{j} \hat{G}^{\mathrm{A}}
    - [\hat{v}^{j} \hat{G}^{\mathrm{R}} \hat{v}^{j} \hat{G}^{\mathrm{R}} + (\mathrm{R} \rightarrow \mathrm{A})]
  \}. \label{eq:conductivity_1}
\end{equation}
The velocity VC~\eqref{eq:velocity_1} in the Fermi-surface term has been taken into account,
and the Fermi-sea term vanishes owing to the cyclic invariance of trace.
Similarly, the SH conductivity is expressed as
\begin{align}
  \sigma_{sa}^{\phantom{sa} ij}
  = & -\frac{\hbar q}{2} \int \frac{d \epsilon}{2 \pi} f(\epsilon) \int \frac{d^{d} k}{(2 \pi)^{d}}
  \tr [
    \hat{J}_{sa}^{\phantom{sa} i} \partial_{\epsilon} \hat{G}^{\mathrm{R}} \hat{v}^{j} \hat{G}^{\mathrm{R}}
    - \hat{J}_{sa}^{\phantom{sa} i} \hat{G}^{\mathrm{R}} \hat{v}^{j} \partial_{\epsilon} \hat{G}^{\mathrm{R}}
    - (\mathrm{R} \rightarrow \mathrm{A})
  ] \notag \\
  & - \frac{\hbar q}{2} \int \frac{d \epsilon}{2 \pi} f^{\prime}(\epsilon) \int \frac{d^{d} k}{(2 \pi)^{d}}
  \tr \{
    2 \hat{J}_{sa}^{\phantom{sa} i} \hat{G}^{\mathrm{R}} \hat{V}^{j} \hat{G}^{\mathrm{A}}
    - [\hat{J}_{sa}^{\phantom{sa} i} \hat{G}^{\mathrm{R}} \hat{v}^{j} \hat{G}^{\mathrm{R}} + (\mathrm{R} \rightarrow \mathrm{A})]
  \}. \label{eq:spin_hall_1}
  \end{align}
  The SAC, namely, the linear response of spin to an electric field gradient, is expressed as
  \begin{align}
    g_{sa}^{\phantom{sa} ij}
    = & \frac{i \hbar q}{4} \int \frac{d \epsilon}{2 \pi} f^{\prime}(\epsilon) \int \frac{d^{d} k}{(2 \pi)^{d}} \notag \\
    & \times \tr \{
      2 (
        \hat{S}_{a} \partial_{k_{i}} \hat{G}^{\mathrm{R}} \hat{V}^{j} \hat{G}^{\mathrm{A}}
        - \hat{S}_{a} \hat{G}^{\mathrm{R}} \hat{V}^{j} \partial_{k_{i}} \hat{G}^{\mathrm{A}}
      ) - [
        \hat{s}_{a} \partial_{k_{i}} \hat{G}^{\mathrm{R}} \hat{v}^{j} \hat{G}^{\mathrm{R}}
        - \hat{s}_{a} \hat{G}^{\mathrm{R}} \hat{v}^{j} \partial_{k_{i}} \hat{G}^{\mathrm{R}}
        + (\mathrm{R} \rightarrow \mathrm{A})
      ]
    \}, \label{spin_accumulation_1}
  \end{align}
  which is schematically shown in Fig.~\ref{fig:diagram}(c).
  Note that the Fermi-sea term can be nonzero only when the time-reversal symmetry is broken.
  The Feynman diagrams as shown in Fig.~\ref{fig:diagram}(d) contribute to the SAC in general
  but not in the current case of short-range impurities.

\section{Application to Dirac model} \label{app:dirac}
We consider the Dirac Hamiltonian,
\begin{equation}
  \hat{H}(\bm{k})
  = -\hbar v \bm{k} \cdot \rho_{y} \bm{\sigma} + \Delta \rho_{z}. \label{eq:dirac}
\end{equation}
Here, $\rho$'s and $\sigma$'s are the Pauli matrices for the orbital and spin degrees of freedom, respectively.
We consider short-range nonmagnetic impurities following Ref.~\cite{JPSJ.86.094704} and denote its magnitude as $v_{\mathrm{i}}$ without hat.

We assume that the self-energy takes the form of
$\hat{\Sigma}^{\mathrm{R}}(\epsilon) = \Sigma^{\mathrm{R} 00}(\epsilon) + \Sigma^{\mathrm{R} z0}(\epsilon) \rho_{z}$.
The retarded Green's function~\eqref{eq:green_1} is expressed as
\begin{equation}
  \hat{G}^{\mathrm{R}}(\epsilon, \bm{k})
  = -\frac{1}{2 x^{\mathrm{R}}(\epsilon)} \sum_{\rho} \frac{\rho}{\hbar v k - \rho \Delta x^{\mathrm{R}}(\epsilon)}
  [g_{+}^{\mathrm{R}}(\epsilon) + g_{-}^{\mathrm{R}}(\epsilon) \rho_{z} - \rho x^{\mathrm{R}}(\epsilon) \hb{k} \cdot \rho_{y} \bm{\sigma}],
  \label{eq:green_3}
\end{equation}
in which
$g_{+}^{\mathrm{R}}(\epsilon) = [\epsilon - \Sigma^{\mathrm{R} 00}(\epsilon)] / \Delta$,
$g_{-}^{\mathrm{R}}(\epsilon) = [\Delta + \Sigma^{\mathrm{R} z0}(\epsilon)] / \Delta$,
$x^{\mathrm{R}}(\epsilon) = \sqrt{[g_{+}^{\mathrm{R}}(\epsilon)]^{2} - [g_{-}^{\mathrm{R}}(\epsilon)]^{2}}$,
and $\hb{k} = \bm{k} / k$ is the normalized wave number.
Hence, the integrated Green's function~\eqref{eq:green_2} is explicitly given by
\begin{equation}
  \hc{G}^{\mathrm{R}}(\epsilon)
  = -\frac{\Delta^{2}}{(\hbar v)^{3}} [I_{0} + I_{1}^{\mathrm{R}}(\epsilon)]
  [g_{+}^{\mathrm{R}}(\epsilon) + g_{-}^{\mathrm{R}}(\epsilon) \rho_{z}]
  \equiv \mathcal{G}^{\mathrm{R} 00}(\epsilon) + \mathcal{G}^{\mathrm{R} z0}(\epsilon) \rho_{z}, \label{eq:green_4}
\end{equation}
in which $I_{0}$ and $I_{1}^{\mathrm{R}}(\epsilon)$ are defined in Eq.~\eqref{eq:integrals}.
The $T$-matrix is calculated as
\begin{equation}
  \hat{T}^{\mathrm{R}}(\epsilon)
  = v_{\mathrm{i}} \frac{
    [1 - v_{\mathrm{i}} \mathcal{G}^{\mathrm{R} 00}(\epsilon)] + v_{\mathrm{i}} \mathcal{G}^{\mathrm{R} z0}(\epsilon) \rho_{z}
  }{
    [1 - v_{\mathrm{i}} \mathcal{G}^{\mathrm{R} 00}(\epsilon)]^{2} - [v_{\mathrm{i}} \mathcal{G}^{\mathrm{R} z0}(\epsilon)]^{2}
  }
  \equiv T^{\mathrm{R} 00}(\epsilon) + T^{\mathrm{R} z0}(\epsilon) \rho_{z}. \label{eq:t-matrix_4}
\end{equation}
Thus, $\hat{\Sigma}^{\mathrm{R}}(\epsilon)$, $\hc{G}^{\mathrm{R}}(\epsilon)$, and $\hat{T}^{\mathrm{R}}(\epsilon)$
are self-consistently determined from the above assumption.
Using the explicit form of Eq.~\eqref{eq:green_4}, the self-energy has two peaks at
$\epsilon / \Delta = -2 \pi^{2} (\hbar v)^{3} / \Delta^{2} v_{\mathrm{i}} \Lambda \pm 1$ with the dimensionless cutoff $\Lambda$ for $\bm{k}$.

Next, we calculate the VC~\eqref{eq:velocity_2} to the velocity operator $\hat{v}^{y} = -v \rho_{y} \sigma_{y}$.
Assuming $\hat{V}^{y}(\epsilon) = V^{yxy}(\epsilon) \rho_{x} \sigma_{y} + V^{yyy}(\epsilon) \rho_{y} \sigma_{y}$,
Eq.~\eqref{eq:velocity_3} takes the same form as
\begin{align}
  \hc{V}^{y}(\epsilon)
  = & \frac{\Delta}{(\hbar v)^{3}} \{
    V^{yxy}(\epsilon) [\xi_{z}(\epsilon) J_{1}(\epsilon) + J_{2}(\epsilon)] - V^{yyy}(\epsilon) \xi_{y}(\epsilon) J_{1}(\epsilon)
  \} \rho_{x} \sigma_{y} \notag \\
  & + \frac{\Delta}{(\hbar v)^{3}} \{
    V^{yxy}(\epsilon) \xi_{y}(\epsilon) J_{1}(\epsilon) + V^{yyy}(\epsilon) [\xi_{z}(\epsilon) J_{1}(\epsilon) - J_{2}(\epsilon)]
  \} \rho_{y} \sigma_{y} \notag \\
  = & \mathcal{V}^{yxy}(\epsilon) \rho_{x} \sigma_{y} + \mathcal{V}^{yyy}(\epsilon) \rho_{y} \sigma_{y}, \label{eq:velocity_4}
\end{align}
in which $J_{1}(\epsilon)$, $J_{2}(\epsilon)$, $\xi_{y}(\epsilon)$, and $\xi_{z}(\epsilon)$ are defined in Eq.~\eqref{eq:functions}.
Thus, $V^{yxy}(\epsilon)$ and $V^{yyy}(\epsilon)$ can be obtained by solving
\begin{equation}
  \begin{bmatrix}
    V^{yxy}(\epsilon) \\
    V^{yyy}(\epsilon)
  \end{bmatrix}
  = \begin{bmatrix}
    0 \\
    -v
  \end{bmatrix}
  + n_{\mathrm{i}} \begin{bmatrix}
    \mathcal{T}_{z}(\epsilon) & -\mathcal{T}_{y}(\epsilon) \\
    \mathcal{T}_{y}(\epsilon) & \mathcal{T}_{z}(\epsilon)
  \end{bmatrix}
  \begin{bmatrix}
    \mathcal{V}^{yxy}(\epsilon) \\
    \mathcal{V}^{yyy}(\epsilon)
  \end{bmatrix}. \label{eq:velocity_5}
\end{equation}
$\mathcal{T}_{y}(\epsilon)$ and $\mathcal{T}_{z}(\epsilon)$ are defined in Eq.~\eqref{eq:t-matrix_6}.
The VC~\eqref{eq:spin_1} to the spin operator $\hat{s}_{z} = (\hbar / 2) \sigma_{z}$ is also available.
We assume $\hat{S}_{z}(\epsilon) = S_{z}^{0z}(\epsilon) \sigma_{z} + S_{z}^{zz}(\epsilon) \rho_{z} \sigma_{z}$,
and Eq.~\eqref{eq:spin_2} takes the same form as
\begin{align}
  \hc{S}_{z}(\epsilon)
  = & \frac{\Delta}{(\hbar v)^{3}} \{
    S_{z}^{0z}(\epsilon) [\xi_{0}(\epsilon) J_{1}(\epsilon) - J_{2}(\epsilon)] + S_{z}^{zz}(\epsilon) \xi_{x}(\epsilon) J_{1}(\epsilon)
  \} \sigma_{z} \notag \\
  & + \frac{\Delta}{(\hbar v)^{3}} \{
    S_{z}^{0z}(\epsilon) \xi_{x}(\epsilon) J_{1}(\epsilon) + S_{z}^{zz}(\epsilon) [\xi_{0}(\epsilon) J_{1}(\epsilon) + J_{2}(\epsilon)]
  \} \rho_{z} \sigma_{z} \notag \\
  = & \mathcal{S}_{z}^{0z}(\epsilon) \sigma_{z} + \mathcal{S}_{z}^{zz}(\epsilon) \rho_{z} \sigma_{z}, \label{eq:spin_3}
\end{align}
with $\xi_{0}(\epsilon)$ and $\xi_{x}(\epsilon)$ in Eq.~\eqref{eq:functions}.
$S_{z}^{0z}(\epsilon)$ and $S_{z}^{zz}(\epsilon)$ are obtained by solving
\begin{equation}
  \begin{bmatrix}
    S_{z}^{0z}(\epsilon) \\
    S_{z}^{zz}(\epsilon)
  \end{bmatrix}
  = \begin{bmatrix}
    \hbar / 2 \\
    0
  \end{bmatrix}
  + n_{\mathrm{i}} \begin{bmatrix}
    \mathcal{T}_{0}(\epsilon) & \mathcal{T}_{x}(\epsilon) \\
    \mathcal{T}_{x}(\epsilon) & \mathcal{T}_{0}(\epsilon)
  \end{bmatrix}
  \begin{bmatrix}
    \mathcal{S}_{z}^{0z}(\epsilon) \\
    \mathcal{S}_{z}^{zz}(\epsilon)
  \end{bmatrix}, \label{eq:spin_4}
\end{equation}
in which $\mathcal{T}_{0}(\epsilon)$ and $\mathcal{T}_{x}(\epsilon)$ are defined in Eq.~\eqref{eq:t-matrix_6}.
The VC $\hat{M}_{z}(\epsilon) = M_{z}^{0z}(\epsilon) \sigma_{z} + M_{z}^{zz}(\epsilon) \rho_{z} \sigma_{z}$
to the magnetic moment operator $\hat{m}_{z} = -(g \mu_{\mathrm{B}} / 2) \rho_{z} \sigma_{z}$ is obtained by solving
the same equations as Eqs.~\eqref{eq:spin_3} and \eqref{eq:spin_4}
with $[\hbar / 2, 0]$ replaced by $[0, -g \mu_{\mathrm{B}} / 2]$.
Aiming at the MMH conductivity, the energy derivative of the self-energy is computed from Eq.~\eqref{eq:t-matrix_3} as
\begin{subequations}\begin{align}
  \begin{bmatrix}
    \mathcal{G}^{\mathrm{R} 00 \prime}(\epsilon) \\
    \mathcal{G}^{\mathrm{R} z0 \prime}(\epsilon)
  \end{bmatrix}
  = & -\frac{\Delta}{2 (\hbar v)^{3}} \begin{bmatrix}
    \zeta_{0}^{\mathrm{R}}(\epsilon) + [2 I_{0} + 3 I_{1}^{\mathrm{R}}(\epsilon) + I_{2}^{\mathrm{R}}(\epsilon)]
    & \zeta_{x}^{\mathrm{R}}(\epsilon) \\
    \zeta_{x}^{\mathrm{R}}(\epsilon)
    & \zeta_{0}^{\mathrm{R}}(\epsilon) - [2 I_{0} + 3 I_{1}^{\mathrm{R}}(\epsilon) + I_{2}^{\mathrm{R}}(\epsilon)]
  \end{bmatrix}
  \begin{bmatrix}
    1 - \Sigma^{\mathrm{R} 00 \prime}(\epsilon) \\
    -\Sigma^{\mathrm{R} z0 \prime}(\epsilon)
  \end{bmatrix}, \label{eq:t-matrix_5a} \\
  \begin{bmatrix}
    T^{\mathrm{R} 00 \prime}(\epsilon) \\
    T^{\mathrm{R} z0 \prime}(\epsilon)
  \end{bmatrix}
  = & \begin{bmatrix}
    \mathcal{T}_{0}^{\mathrm{R}}(\epsilon) & \mathcal{T}_{x}^{\mathrm{R}}(\epsilon) \\
    \mathcal{T}_{x}^{\mathrm{R}}(\epsilon) & \mathcal{T}_{0}^{\mathrm{R}}(\epsilon)
  \end{bmatrix}
  \begin{bmatrix}
    \mathcal{G}^{\mathrm{R} 00 \prime}(\epsilon) \\
    \mathcal{G}^{\mathrm{R} z0 \prime}(\epsilon)
  \end{bmatrix},
\end{align}\label{eq:t-matrix_5}\end{subequations}
in which $I_{2}^{\mathrm{R}}(\epsilon)$ is defined in Eq.~\eqref{eq:integrals_3},
$\zeta_{0}^{\mathrm{R}}(\epsilon)$ and $\zeta_{x}^{\mathrm{R}}(\epsilon)$ are in Eq.~\eqref{eq:functions},
and $\mathcal{T}_{0}^{\mathrm{R}}(\epsilon)$ and $\mathcal{T}_{x}^{\mathrm{R}}(\epsilon)$ are in Eq.~\eqref{eq:t-matrix_6}.

Finally, the electric conductivity, SAC, MMAC, and MMH conductivity are computed as
\begin{equation}
  \sigma^{yy}
  = \frac{2 q^{2} \Delta}{\pi (\hbar v)^{2}} \int d \epsilon f^{\prime}(\epsilon) \left\{
    V^{yxy}(\epsilon) \xi_{y}(\epsilon) J_{1}(\epsilon) + V^{yyy}(\epsilon) [\xi_{z}(\epsilon) J_{1}(\epsilon) - J_{2}(\epsilon)]
    - \frac{v}{6} [2 I_{0} - I_{2}^{\mathrm{R}}(\epsilon) - I_{2}^{\mathrm{A}}(\epsilon)]
  \right\}, \label{eq:electric_conductivity}
\end{equation}
\begin{align}
  g_{sz}^{\phantom{sz} xy}
  = & \frac{q}{\pi \hbar v^{2}} \int d \epsilon f^{\prime}(\epsilon) \left(
    S_{z}^{0z}(\epsilon) [V^{yxy}(\epsilon) K_{++}(\epsilon) + V^{yyy}(\epsilon) K_{+-}(\epsilon)]
    + S_{z}^{zz}(\epsilon) [V^{yxy}(\epsilon) K_{-+}(\epsilon) + V^{yyy}(\epsilon) K_{--}(\epsilon)]
  \right. \notag \\
  & \left.
    + \frac{\hbar v}{4} \left\{
      \frac{g_{+}^{\mathrm{R}}(\epsilon) [I_{1}^{\mathrm{R}}(\epsilon) + I_{2}^{\mathrm{R}}(\epsilon)]}{[x^{\mathrm{R}}(\epsilon)]^{2}}
      + (\mathrm{R} \rightarrow \mathrm{A})
    \right\}
  \right), \label{eq:spin_accumulation}
\end{align}
\begin{align}
  g_{mz}^{\phantom{mz} xy}
  = & \frac{q}{\pi \hbar v^{2}} \int d \epsilon f^{\prime}(\epsilon) \left(
    M_{z}^{0z}(\epsilon) [V^{yxy}(\epsilon) K_{++}(\epsilon) + V^{yyy}(\epsilon) K_{+-}(\epsilon)]
    + M_{z}^{zz}(\epsilon) [V^{yxy}(\epsilon) K_{-+}(\epsilon) + V^{yyy}(\epsilon) K_{--}(\epsilon)]
  \right. \notag \\
  & \left.
    - \frac{g \mu_{\mathrm{B}} v}{4} \left\{
      \frac{g_{-}^{\mathrm{R}}(\epsilon) [I_{1}^{\mathrm{R}}(\epsilon) + I_{2}^{\mathrm{R}}(\epsilon)]}{[x^{\mathrm{R}}(\epsilon)]^{2}}
      + (\mathrm{R} \rightarrow \mathrm{A})
    \right\}
  \right), \label{eq:magnetic_moment_accumulation}
\end{align}
\begin{align}
  \sigma_{mz}^{\phantom{mz} xy}
  = & \frac{g \mu_{\mathrm{B}} q}{2 \pi \hbar^{2} v} \int d \epsilon f(\epsilon) \left(
    i \frac{
      \{
        [1 - \Sigma^{\mathrm{R} 00 \prime}(\epsilon)] g_{-}^{\mathrm{R}}(\epsilon)
        - \Sigma^{\mathrm{R} z0 \prime}(\epsilon) g_{+}^{\mathrm{R}}(\epsilon)
      \} [I_{1}^{\mathrm{R}}(\epsilon) + I_{2}^{\mathrm{R}}(\epsilon)]
    }{[x^{\mathrm{R}}(\epsilon)]^{2}}
    - (\mathrm{R} \rightarrow \mathrm{A})
  \right) \notag \\
  & - \frac{g \mu_{\mathrm{B}} q \Delta}{\pi (\hbar v)^{2}} \int d \epsilon f^{\prime}(\epsilon)
  \{V^{yxy}(\epsilon) [\xi_{z}(\epsilon) J_{1}(\epsilon) + J_{2}(\epsilon)] - V^{yyy}(\epsilon) \xi_{y}(\epsilon) J_{1}(\epsilon)\}.
  \label{eq:magnetic_moment_hall_conductivity}
\end{align}
$K(\epsilon)$'s are given in Eq.~\eqref{eq:integrals}.

In the above, we have defined dimensionless integrals as
\begin{subequations}\begin{align}
  I_{0}
  = & \frac{1}{4 \pi^{2}} \sum_{\rho} \int_{0}^{\Lambda} d x
  = \frac{\Lambda}{2 \pi^{2}}, \label{eq:integrals_1} \\
  I_{1}^{\mathrm{R}}(\epsilon)
  = & \frac{1}{4 \pi^{2}} \sum_{\rho} \int_{0}^{\Lambda} d x \frac{\rho x^{\mathrm{R}}(\epsilon)}{x - \rho x^{\mathrm{R}}(\epsilon)}
  = \frac{1}{4 \pi^{2}} x^{\mathrm{R}}(\epsilon)
  \ln \frac{x^{\mathrm{R}}(\epsilon) - \Lambda}{x^{\mathrm{R}}(\epsilon) + \Lambda}, \label{eq:integrals_2} \\
  I_{2}^{\mathrm{R}}(\epsilon)
  = & \frac{1}{4 \pi^{2}} \sum_{\rho} \int_{0}^{\Lambda} d x \left[
    \frac{x^{\mathrm{R}}(\epsilon)}{x - \rho x^{\mathrm{R}}(\epsilon)}
  \right]^{2}
  = \frac{1}{2 \pi^{2}} \frac{[x^{\mathrm{R}}(\epsilon)]^{2} \Lambda}{[x^{\mathrm{R}}(\epsilon)]^{2} - \Lambda^{2}}, \label{eq:integrals_3}
\end{align}\label{eq:integrals}\end{subequations}
in which we have changed a variable as $x = \hbar v k / \Delta$ and introduced its cutoff $\Lambda$.
We have also defined dimensionless functions as
\begin{subequations}\begin{align}
  \zeta_{0}^{\mathrm{R}}(\epsilon)
  = & \frac{
    \{[g_{+}^{\mathrm{R}}(\epsilon)]^{2} + [g_{-}^{\mathrm{R}}(\epsilon)]^{2}\}
    [I_{1}^{\mathrm{R}}(\epsilon) + I_{2}^{\mathrm{R}}(\epsilon)]
  }{[x^{\mathrm{R}}(\epsilon)]^{2}}, \label{eq:functions_1} \\
  \zeta_{x}^{\mathrm{R}}(\epsilon)
  = & \frac{
    2 g_{+}^{\mathrm{R}}(\epsilon) g_{-}^{\mathrm{R}}(\epsilon)
    [I_{1}^{\mathrm{R}}(\epsilon) + I_{2}^{\mathrm{R}}(\epsilon)]
  }{[x^{\mathrm{R}}(\epsilon)]^{2}}, \label{eq:functions_2} \\
  \xi_{0}(\epsilon)
  = & g_{+}^{\mathrm{R}}(\epsilon) g_{+}^{\mathrm{A}}(\epsilon) + g_{-}^{\mathrm{R}}(\epsilon) g_{-}^{\mathrm{A}}(\epsilon),
  \label{eq:functions_3} \\
  \xi_{x}(\epsilon)
  = & g_{+}^{\mathrm{R}}(\epsilon) g_{-}^{\mathrm{A}}(\epsilon) + g_{-}^{\mathrm{R}}(\epsilon) g_{+}^{\mathrm{A}}(\epsilon),
  \label{eq:functions_4} \\
  \xi_{y}(\epsilon)
  = & -i [g_{+}^{\mathrm{R}}(\epsilon) g_{-}^{\mathrm{A}}(\epsilon) - g_{-}^{\mathrm{R}}(\epsilon) g_{+}^{\mathrm{A}}(\epsilon)],
  \label{eq:functions_5} \\
  \xi_{z}(\epsilon)
  = & g_{+}^{\mathrm{R}}(\epsilon) g_{+}^{\mathrm{A}}(\epsilon) - g_{-}^{\mathrm{R}}(\epsilon) g_{-}^{\mathrm{A}}(\epsilon),
  \label{eq:functions_6} \\
  J_{1}(\epsilon)
  = & \frac{I_{1}^{\mathrm{R}}(\epsilon) - I_{1}^{\mathrm{A}}(\epsilon)}{[x^{\mathrm{R}}(\epsilon)]^{2} - [x^{\mathrm{A}}(\epsilon)]^{2}},
  \label{eq:functions_7} \\
  J_{2}(\epsilon)
  = & \frac{1}{3} \left\{
    I_{0} + \frac{
      [x^{\mathrm{R}}(\epsilon)]^{2} I_{1}^{\mathrm{R}}(\epsilon) - [x^{\mathrm{A}}(\epsilon)]^{2} I_{1}^{\mathrm{A}}(\epsilon)
    }{[x^{\mathrm{R}}(\epsilon)]^{2} - [x^{\mathrm{A}}(\epsilon)]^{2}}
  \right\}, \label{eq:functions_8} \\
  J_{3}(\epsilon)
  = & \frac{1}{3} \frac{i}{[x^{\mathrm{R}}(\epsilon)]^{2} - [x^{\mathrm{A}}(\epsilon)]^{2}} \left(
    I_{1}^{\mathrm{R}}(\epsilon) + I_{1}^{\mathrm{A}}(\epsilon) + I_{2}^{\mathrm{R}}(\epsilon) + I_{2}^{\mathrm{A}}(\epsilon) - \frac{
      2 \{[x^{\mathrm{R}}(\epsilon)]^{2} + [x^{\mathrm{A}}(\epsilon)]^{2}\} [I_{1}^{\mathrm{R}}(\epsilon) - I_{1}^{\mathrm{A}}(\epsilon)]
    }{[x^{\mathrm{R}}(\epsilon)]^{2} - [x^{\mathrm{A}}(\epsilon)]^{2}}
  \right), \label{eq:functions_9} \\
  K_{++}(\epsilon)
  = & i [g_{-}^{\mathrm{R}}(\epsilon) - g_{-}^{\mathrm{A}}(\epsilon)] J_{1}(\epsilon)
  + [g_{-}^{\mathrm{R}}(\epsilon) + g_{-}^{\mathrm{A}}(\epsilon)] J_{3}(\epsilon), \label{eq:functions_10} \\
  K_{+-}(\epsilon)
  = & [g_{+}^{\mathrm{R}}(\epsilon) + g_{+}^{\mathrm{A}}(\epsilon)] J_{1}(\epsilon)
  - i [g_{+}^{\mathrm{R}}(\epsilon) - g_{+}^{\mathrm{A}}(\epsilon)] J_{3}(\epsilon), \label{eq:functions_11} \\
  K_{-+}(\epsilon)
  = & i [g_{+}^{\mathrm{R}}(\epsilon) - g_{+}^{\mathrm{A}}(\epsilon)] J_{1}(\epsilon)
  + [g_{+}^{\mathrm{R}}(\epsilon) + g_{+}^{\mathrm{A}}(\epsilon)] J_{3}(\epsilon), \label{eq:functions_12} \\
  K_{--}(\epsilon)
  = & [g_{-}^{\mathrm{R}}(\epsilon) + g_{-}^{\mathrm{A}}(\epsilon)] J_{1}(\epsilon)
  - i [g_{-}^{\mathrm{R}}(\epsilon) - g_{-}^{\mathrm{A}}(\epsilon)] J_{3}(\epsilon). \label{eq:functions_13}
\end{align}\label{eq:functions}\end{subequations}
Also, we have introduced
\begin{subequations}\begin{align}
  \mathcal{T}_{0}^{\mathrm{R}}(\epsilon)
  = & [T^{\mathrm{R} 00}(\epsilon)]^{2} + [T^{\mathrm{R} z0}(\epsilon)]^{2}, \label{eq:t-matrix_6a} \\
  \mathcal{T}_{x}^{\mathrm{R}}(\epsilon)
  = & 2 T^{\mathrm{R} 00}(\epsilon) T^{\mathrm{R} z0}(\epsilon), \label{eq:t-matrix_6b} \\
  \mathcal{T}_{0}(\epsilon)
  = & T^{\mathrm{R} 00}(\epsilon) T^{\mathrm{A} 00}(\epsilon)
  + T^{\mathrm{R} z0}(\epsilon) T^{\mathrm{A} z0}(\epsilon), \label{eq:t-matrix_6c} \\
  \mathcal{T}_{x}(\epsilon)
  = & T^{\mathrm{R} 00}(\epsilon) T^{\mathrm{A} z0}(\epsilon)
  + T^{\mathrm{A} 00}(\epsilon) T^{\mathrm{R} z0}(\epsilon), \label{eq:t-matrix_6d} \\
  \mathcal{T}_{y}(\epsilon)
  = & -i [T^{\mathrm{R} 00}(\epsilon) T^{\mathrm{A} z0}(\epsilon)
  - T^{\mathrm{A} 00}(\epsilon) T^{\mathrm{R} z0}(\epsilon)], \label{eq:t-matrix_6e} \\
\mathcal{T}_{z}(\epsilon)
  = & T^{\mathrm{R} 00}(\epsilon) T^{\mathrm{A} 00}(\epsilon)
  - T^{\mathrm{R} z0}(\epsilon) T^{\mathrm{A} z0}(\epsilon). \label{eq:t-matrix_6f}
\end{align}\label{eq:t-matrix_6}\end{subequations}

\section{Bloch formula} \label{app:bloch}
We provide the SAC and MMAC within the relaxation time approximation, which are expressed by Bloch wave functions.
The Dirac Hamiltonian~\eqref{eq:dirac} commutes with the helicity operator $\hat{\eta}(\hb{k}) = \hb{k} \cdot \bm{\sigma}$
and hence has simultaneous wave functions.
Hereafter, $\hat{\eta}(\hb{k})$ may act on a two- or four-component spinor.
The eigenvalues are $\epsilon_{\lambda}(k) = \lambda \epsilon(k) = \lambda \sqrt{(\hbar v k)^{2} + \Delta^{2}}$ with $\lambda = \pm 1$,
each of which is doubly degenerate owing to the inversion and time-reversal symmetries.
Two positive-energy wave functions are expressed by
\begin{equation}
  | u_{+, \eta}(\bm{k}) \rangle
  = \begin{bmatrix}
    u(k) \chi_{\eta}(\hb{k}) \\
    -i \eta v(k) \chi_{\eta}(\hb{k})
  \end{bmatrix} \label{eq:positive_wave_function},
\end{equation}
in which $u(k) = \sqrt{[1 + \Delta / \epsilon(k)] / 2}, v(k) = \sqrt{[1 - \Delta / \epsilon(k)] / 2}$,
and $\eta = \pm 1$ and $\chi_{\eta}(\hb{k})$ are the eigenvalue and two-component wave function of $\hat{\eta}(\hb{k})$,
\begin{align}
  \chi_{+}(\hb{k})
  = & \begin{bmatrix}
    \cos \theta / 2 \\
    e^{i \phi} \sin \theta / 2
  \end{bmatrix}, &
  \chi_{-}(\hb{k})
  = & \begin{bmatrix}
    -e^{-i \phi} \sin \theta / 2 \\
    \cos \theta / 2
  \end{bmatrix}. \label{eq:helicity}
\end{align}
Also, this system has the particle-hole symmetry $\hat{C} = \rho_{y} \sigma_{y} \hat{K}$ with the complex conjugate operator $\hat{K}$.
If $| u(\bm{k}) \rangle$ is an eigenvector of $\hat{H}(\bm{k})$ with the eigenvalue $\epsilon(\bm{k})$,
$\hat{C} | u(-\bm{k}) \rangle$ is an eigenvector of $\hat{H}(\bm{k})$ with the eigenvalue $-\epsilon(-\bm{k})$.
Thus, two negative-energy wave functions are expressed by $| u_{-, \eta}(\bm{k}) \rangle = \hat{C} | u_{+, \eta}(-\bm{k}) \rangle$.

We compute the SAC and MMAC within the relaxation time approximation.
Since the previous formula~\cite{PhysRevB.105.L201202} cannot be applied to degenerate cases, we modify it as
\begin{subequations}\begin{align}
  g_{sz}^{\phantom{sz} xy}
  = & -\frac{q \tau}{\hbar} \sum_{\lambda \eta \eta^{\prime}} \int \frac{d^{3} k}{(2 \pi)^{3}}
  \{[s_{\lambda z}^{\phantom{\lambda z} x}(\bm{k})]_{\eta \eta^{\prime}} \partial_{k_{y}} \epsilon_{\lambda}(k) \delta_{\eta^{\prime} \eta}
  - [s_{\lambda z}(\bm{k})]_{\eta \eta^{\prime}} [m_{\lambda z}^{(\mathrm{orb})}(\bm{k})]_{\eta^{\prime} \eta}\}
  [-f^{\prime}(\epsilon_{\lambda}(k))], \label{eq:spin_accumulation_bloch_1} \\
  g_{mz}^{\phantom{sz} xy}
  = & -\frac{q \tau}{\hbar} \sum_{\lambda \eta \eta^{\prime}} \int \frac{d^{3} k}{(2 \pi)^{3}}
  \{[m_{\lambda z}^{\phantom{\lambda z} x}(\bm{k})]_{\eta \eta^{\prime}} \partial_{k_{y}} \epsilon_{\lambda}(k) \delta_{\eta^{\prime} \eta}
  - [m_{\lambda z}(\bm{k})]_{\eta \eta^{\prime}} [m_{\lambda z}^{(\mathrm{orb})}(\bm{k})]_{\eta^{\prime} \eta}\}
  [-f^{\prime}(\epsilon_{\lambda}(k))]. \label{eq:magnetic-moment_accumulation_bloch_1}
\end{align}\label{eq:bloch_1}\end{subequations}
Here, the matrix elements of spin and magnetic moment are computed as
\begin{subequations}\begin{align}
  [s_{\lambda z}(\bm{k})]_{\eta \eta^{\prime}}
  = & \langle u_{\lambda \eta}(\bm{k}) | \hat{s}_{z} | u_{\lambda \eta^{\prime}}(\bm{k}) \rangle
  = \frac{\hbar}{2 \epsilon(k)} \begin{bmatrix}
    \epsilon(k) \cos \theta & -e^{-i \lambda \phi} \Delta \sin \theta \\
    -e^{i \lambda \phi} \Delta \sin \theta & -\epsilon(k) \cos \theta
  \end{bmatrix}, \label{eq:spin_matrix_elements} \\
  [m_{\lambda z}(\bm{k})]_{\eta \eta^{\prime}}
  = & \langle u_{\lambda \eta}(\bm{k}) | \hat{m}_{z} | u_{\lambda \eta^{\prime}}(\bm{k}) \rangle
  = -\lambda \frac{g \mu_{\mathrm{B}}}{2 \epsilon(k)} \begin{bmatrix}
    \Delta \cos \theta & -e^{-i \lambda \phi} \epsilon(k) \sin \theta \\
    -e^{i \lambda \phi} \epsilon(k) \sin \theta & -\Delta \cos \theta
  \end{bmatrix}, \label{eq:magneic-moment_matrix_elements}
\end{align}\end{subequations}
their quadrupole moments~\cite{PhysRevB.97.134423,PhysRevB.99.024404} are expressed by
\begin{subequations}\begin{align}
  [s_{\lambda z}^{\phantom{\lambda z} x}(\bm{k})]_{\eta \eta^{\prime}}
  = & \frac{i}{2} \sum_{\kappa (\not= \lambda) \xi} \frac{
    \langle u_{\lambda \eta}(\bm{k}) | \hat{s}_{z} | u_{\kappa \xi}(\bm{k}) \rangle
    \langle u_{\kappa \xi}(\bm{k}) | \hbar \hat{v}^{x}(\bm{k}) | u_{\lambda \eta^{\prime}}(\bm{k}) \rangle
    - \langle u_{\lambda \eta}(\bm{k}) | \hbar \hat{v}^{x}(\bm{k}) | u_{\kappa \xi}(\bm{k}) \rangle
    \langle u_{\kappa \xi}(\bm{k}) | \hat{s}_{z} | u_{\lambda \eta^{\prime}}(\bm{k}) \rangle
    }{
    \epsilon_{\lambda}(k) - \epsilon_{\kappa}(k)
    } \notag \\
  = & -\frac{\hbar^{3} v^{2} k \sin \theta \sin \phi}{4 [\epsilon(k)]^{2}} \begin{bmatrix}
    1 & 0 \\
    0 & 1
  \end{bmatrix}, \label{eq:spin_quadrupole} \\
  [m_{\lambda z}^{\phantom{\lambda z} x}(\bm{k})]_{\eta \eta^{\prime}}
  = & \frac{i}{2} \sum_{\kappa (\not= \lambda) \xi} \frac{
    \langle u_{\lambda \eta}(\bm{k}) | \hat{m}_{z} | u_{\kappa \xi}(\bm{k}) \rangle
    \langle u_{\kappa \xi}(\bm{k}) | \hbar \hat{v}^{x}(\bm{k}) | u_{\lambda \eta^{\prime}}(\bm{k}) \rangle
    - \langle u_{\lambda \eta}(\bm{k}) | \hbar \hat{v}^{x}(\bm{k}) | u_{\kappa \xi}(\bm{k}) \rangle
    \langle u_{\kappa \xi}(\bm{k}) | \hat{m}_{z} | u_{\lambda \eta^{\prime}}(\bm{k}) \rangle
    }{
    \epsilon_{\lambda}(k) - \epsilon_{\kappa}(k)
    } \notag \\
  = & \begin{bmatrix}
    0 & 0 \\
    0 & 0
  \end{bmatrix}, \label{eq:magnetic-moment_quadrupole}
\end{align}\label{eq:quadrupole}\end{subequations}
and the group velocity and the orbital magnetic moment are expresed by
\begin{subequations}\begin{align}
  \partial_{k_{y}} \epsilon_{\lambda}(k)
  = & \lambda \frac{\hbar^{2} v^{2} k \sin \theta \sin \phi}{\epsilon(k)}, \label{eq:group_velocity} \\
  [m_{\lambda z}^{(\mathrm{orb})}(\bm{k})]_{\eta \eta^{\prime}}
  = & -\frac{i}{2} \sum_{\kappa (\not= \lambda) \xi} \frac{
    \langle u_{\lambda \eta}(\bm{k}) | \hbar \hat{v}^{x}(\bm{k}) | u_{\kappa \xi}(\bm{k}) \rangle
    \langle u_{\kappa \xi}(\bm{k}) | \hbar \hat{v}^{y}(\bm{k}) | u_{\lambda \eta^{\prime}}(\bm{k}) \rangle - (x \leftrightarrow y)
  }{
    \epsilon_{\lambda}(\bm{k}) - \epsilon_{\kappa}(\bm{k})
  } \notag \\
  = & \lambda \frac{(\hbar v)^{2}}{2 [\epsilon(k)]^{2}} \begin{bmatrix}
    \epsilon(k) \cos \theta & -e^{-i \lambda \phi} \Delta \sin \theta \\
    -e^{i \lambda \phi} \Delta \sin \theta & -\epsilon(k) \cos \theta
  \end{bmatrix}. \label{eq:orbital_magnetic_moment}
\end{align}\label{eq:orbital}\end{subequations}
Thus, at $\mu = \epsilon_{\mathrm{F}}$ and $T = 0$, Eq.~\eqref{eq:bloch_1} is reduced to
\begin{subequations}\begin{align}
  g_{sz}^{\phantom{sz} xy}
  = & -\frac{q \tau}{\hbar} \sum_{\lambda \eta} \int \frac{d^{3} k}{(2 \pi)^{3}}
  (-\lambda) \frac{\hbar^{3} v^{2}}{4 [\epsilon(k)]^{3}}
  \{(\hbar v k)^{2} \sin^{2} \theta \sin^{2} \phi + [\epsilon(k)]^{2} \cos^{2} \theta + \Delta^{2} \sin^{2} \theta\}
  [-f^{\prime}(\epsilon_{\lambda}(k))] \notag \\
  = & \frac{q \tau \Delta}{12 \pi^{2} \hbar v} \sum_{\lambda} \lambda \int_{0}^{\infty} d x
  \frac{x^{2} (2 x^{2} + 3)}{(x^{2} + 1)^{3 / 2}} \delta(\sqrt{x^{2} + 1} - \lambda \epsilon_{\mathrm{F}} / \Delta) \notag \\
  = & \frac{q \tau}{12 \pi^{2} \hbar v} \sgn \epsilon_{\mathrm{F}}
  \frac{\sqrt{\epsilon_{\mathrm{F}}^{2} - \Delta^{2}} (2 \epsilon_{\mathrm{F}}^{2} + \Delta^{2})}{\epsilon_{\mathrm{F}}^{2}},
  \label{eq:spin_accumulation_bloch_2} \\
  g_{mz}^{\phantom{mz} xy}
  = & -\frac{q \tau}{\hbar} \sum_{\lambda \eta} \int \frac{d^{3} k}{(2 \pi)^{3}}
  \frac{g \mu_{\mathrm{B}} \hbar^{2} v^{2} \Delta}{4 [\epsilon(k)]^{2}}
  [-f^{\prime}(\epsilon_{\lambda}(k))] \notag \\
  = & -\frac{g \mu_{\mathrm{B}} q \tau \Delta}{4 \pi^{2} \hbar^{2} v} \sum_{\lambda} \int_{0}^{\infty} d x
  \frac{x^{2}}{x^{2} + 1} \delta(\sqrt{x^{2} + 1} - \lambda \epsilon_{\mathrm{F}} / \Delta) \notag \\
  = & -\frac{g \mu_{\mathrm{B}} q \tau \Delta}{4 \pi^{2} \hbar^{2} v}
  \frac{\sqrt{\epsilon_{\mathrm{F}}^{2} - \Delta^{2}}}{|\epsilon_{\mathrm{F}}|}. \label{eq:magnetic_moment_accumulation_bloch_2}
\end{align}\label{eq:bloch_2}\end{subequations}
These results coincide with the numerical ones by assuming the scalar self-energy and neglecting the VCs.
\end{widetext}
\end{document}